\documentclass[journal]{IEEEtran}
\usepackage{amsmath,amssymb,amsfonts}
\usepackage{algorithmic}

\usepackage[linesnumbered, ruled]{algorithm2e}
\usepackage{array}
\usepackage{booktabs}
\usepackage{textcomp}
\usepackage{stfloats}
\usepackage{url}
\usepackage{verbatim}
\usepackage{graphicx}
\usepackage{cite}
\usepackage[capitalize]{cleveref}
\usepackage[caption=false]{subfig} 
\usepackage{float}
\usepackage{multicol}
\usepackage{multirow}

\usepackage{epstopdf}

\newenvironment{proof}{{\noindent\it Proof.}\quad}{\hfill $\square$\par}

\begin{document}

\title{DFedSat: Communication-Efficient and Robust Decentralized Federated Learning for \\ LEO Satellite Constellations}

\author{Minghao Yang, Jingjing Zhang, Shengyun Liu
\thanks{Minghao Yang and Jingjing Zhang are with the Department of Communication
Science and Engineering, Fudan University, Shanghai, China. Shengyun Liu is with the School of Electronic, Information and Electrical Engineering, Shanghai Jiao Tong University, Shanghai, China (e-mail:
23210720070@m.fudan.edu.cn, jingjingzhang@fudan.edu.cn, shengyun.liu@sjtu.edu.cn). (\textit{Corresponding author: Jingjing Zhang)}}
}

\maketitle

\begin{abstract}
Low Earth Orbit (LEO) satellites play a crucial role in the development of 6G mobile networks and space-air-ground integrated systems. Recent advancements in space technology have empowered LEO satellites with the capability to run AI applications. However, centralized approaches, where ground stations (GSs) act as servers and satellites as clients, often encounter slow convergence and inefficiencies due to intermittent connectivity between satellites and GSs. In contrast, decentralized federated learning (DFL) offers a promising alternative by facilitating direct communication between satellites (clients) via inter-satellite links (ISLs).
However, inter-plane ISLs connecting satellites from different orbital planes are dynamic due to Doppler shifts and pointing limitations. This could impact model propagation and lead to slower convergence. To mitigate these issues, we propose DFedSat, a fully decentralized federated learning framework tailored for LEO satellites. 
DFedSat accelerates the training process by employing two adaptive mechanisms for intra-plane and inter-plane model aggregation, respectively. Furthermore, a self-compensation mechanism is integrated to enhance the robustness of inter-plane ISLs against transmission failure. Additionally, we derive the sublinear convergence rate for the non-convex case of DFedSat. Extensive experimental results demonstrate DFedSat's superiority over other DFL baselines regarding convergence rate, communication efficiency, and resilience to unreliable links. 
\end{abstract}


\section{Introduction}
Satellite communication technology is poised to become a key component of the future integrated air, space, and ground network, attracting significant interest from industry giants such as SpaceX and Amazon, as well as governmental organizations like ESA and NASA \cite{9473799}. This enthusiasm has driven the deployment of numerous Low Earth Orbit (LEO) satellites. Advancements in satellite hardware have empowered LEO satellites with enhanced cameras, processors, and antennas, enabling the collection of extensive Earth imagery and sensor data \cite{harris2018tech,xie2020satellite}. Leveraging machine learning (ML) techniques to analyze these large datasets can provide valuable insights for a wide range of applications, including urban planning, weather forecasting, climate change research, and disaster management \cite{Perez-Portero_Munoz-Martin_Park_Camps_2021,sensing,mateo2021towards}.

However, the conventional methodology of transferring raw image data to a central server, such as a ground station (GS), for training a centralized ML model is impractical due to limited bandwidth and data privacy concerns. Federated learning (FL) offers a promising solution to these challenges \cite{McMahan_Moore_Ramage_Hampson_Arcas_2016}. Within the framework of FL, each client (satellite) independently updates ML models based on its local dataset, transmitting only model parameters in each iteration. This approach can significantly reduce bandwidth requirements while preserving data privacy.

Although research on satellite FL is still in its early stages, several studies have made significant progress. The traditional synchronous FL method, FedAvg \cite{McMahan_Moore_Ramage_Hampson_Arcas_2016}, is applied to LEO constellations, demonstrating the benefits of combining satellite networks and FL \cite{Chen_Xiao_Pang_2022}. However, the short, sporadic, and irregular communication windows between GS and satellites result in very slow convergence. To enhance the performance in satellite FL systems, researchers apply ISLs and clustering strategies \cite{razmi2022board,lin2022federated,chen2023edge,elmahallawy2023optimizing}. Additionally, High-Altitude Platforms (HAPs) are introduced as relays to facilitate faster convergence \cite{fedhap,elmahallawy2024communication}. To circumvent the limitations imposed by communication windows, asynchronous FL is applied to LEO constellations \cite{razmi2022ground}. However, asynchronous FL encounters the issue of gradient staleness, where outdated gradients can significantly impair model convergence and overall performance \cite{xie2019asynchronous}. 
Recent works attempt to mitigate this impact through rational scheduling \cite{razmi2022scheduling}, adjusting the weights of the outdated gradients\cite{so2022fedspace,elmahallawy2022asyncfleo}, optimizing the transmission delay to reduce the outdated gradients\cite{wang2022fl}, and employing gradient compensation \cite{FedGSM}.



In contrast to centralized federated learning (CFL), decentralized federated learning (DFL) overcomes the limitations of the parameter-server structure
\cite{Lian_Zhang_Zhang_Hsieh_Zhang_Liu_2017}. 
The authors of \cite{Sun_Li_Wang_2022} introduce refinements to DFL by employing multiple local iterations and quantization techniques to reduce communication costs, yielding promising convergence results, particularly in convex scenarios. More recent contributions, such as \cite{shi2023improving}, propose methodologies like DFedSAM, which leverages gradient perturbation to generate local flat models via sharpness-aware minimization (SAM). This method achieves optimal results in current DFL-related work. 
In the decentralized satellite FL framework, model parameters are transmitted between satellites through inter-satellite links (ISLs). Research on decentralized satellite FL (DSFL) remains relatively sparse so far. Wu et al. propose DSFL \cite{wu2022dsfl}, which uses intra-plane and inter-plane satellite links to accomplish model aggregation. It designs an inter-satellite routing algorithm to aggregate the model onto a specific satellite. Most recently, \cite{fedleo} proposes an offloading-assisted DSFL framework, aiming to speed up convergence by facilitating data sharing among satellites to counteract the effects of heterogeneous data distribution. However, they assume ideally that the transmission must be 100\% successful.



In real satellite communication networks, inter-plane ISLs are usually unreliable due to two main factors:

$\bullet$ Doppler Shift: The rapid relative movement of satellites induces a significant, time-varying Doppler shift that impairs communication between satellites \cite{doppler1,doppler2,doppler3}.

$\bullet$ Pointing Errors: Misalignment of the transceivers leads to pointing errors, resulting in additional performance degradation and even link failure \cite{pointing1,pointing2,pointing3}.
 


The impact of such unreliability can result in the loss of the received signal, or even slow convergence rates and low communication efficiency. Therefore, for a more realistic study of DSFL, the instability of inter-plane ISLs must be considered. To the best of our knowledge, this is the first work to address this instability in DSFL. This study aims to bridge this gap by considering non-idealized ISLs in the LEO constellation and tailoring the DSFL framework for efficient and robust operation. To this end, we develop a robust and fully decentralized satellite FL framework tailored to LEO satellite networks, namely DFedSat. 
To address the issue of unreliable inter-plane transmission, DFedSat employs a robust model self-compensation mechanism to handle packet loss and transmission errors that may occur during model transmission.

Our main contributions are summarized as follows.

$\bullet$ We present DFedSat, a fully decentralized framework for satellite federated learning. To the best of our knowledge, this is the first work to address the instability of ISLs in decentralized satellite FL. The proposed framework enables distributed consensus of models by implementing two specialized mechanisms tailored for intra-plane and inter-plane model exchanges.

$\bullet$  To mitigate the adverse effects of unreliable inter-plane ISLs on the model transfer process, we introduce a self-compensation mechanism within our framework. This approach not only enhances the robustness of the inter-plane communication but also mitigates the overall communication overhead by reducing the need for retransmissions.

$\bullet$ We present the sublinear convergence rate for DFedSat in a non-convex setting under conditions of unreliable communication. Our findings indicate that DFedSat attains an asymptotic convergence rate comparable to that of the standard DFL algorithm in scenarios with perfect communication. This establishes the robustness and reliability of DFedSat, particularly in the presence of instabilities in inter-plane ISLs.

$\bullet$ We conduct extensive experiments to demonstrate the efficacy of DFedSat. Experimental results show that DFedSat exhibits higher performance and communication efficiency compared to the benchmark strategies. Furthermore, it demonstrates superior performance with high robustness under unreliable communication conditions.


The remainder of this paper is organized as follows. Section II presents the system model for LEO Satellite Constellations and the framework of DFL. Section III elaborates on the proposed DFedSat algorithm, specifically designed for the LEO constellations. Section IV provides a convergence analysis of DFedSat. Numerical results are discussed in Section V, and conclusions are drawn in Section VI.

\textit{Notations}: Italic, bold lower-case, and bold upper-case
letters represent scalars, column vectors, and matrices, respectively. The notation ·\%· represents the modulo operation, i.e., $m\%n$ calculates the remainder
when dividing $m$ by $n$. And the operator $|\cdot|$ indicates
the cardinality of a set or the absolute value of a scalar, while $\|\cdot\|$ denotes the Euclidean norm. The symbol $\odot$ represents the element-wise multiplication. Given a binary vector \(\mathbf{m}\), the negation of \(\mathbf{m}\) is denoted as \(\neg \mathbf{m}\), where each element is flipped from 0 to 1 or from 1 to 0. The operator $(\cdot)^\top$ denotes the transpose of a matrix. Finally, $\mathbb{E}\{\cdot\}$ denotes the expectation.

\section{System Model}

In this section, we introduce the system model of the LEO system. This includes the architecture of the LEO  Constellations, the satellite communication model, the satellite computing model within a DFL framework, and the ISL packet failure model.

\subsection{LEO Satellite Constellations}

\begin{figure}[t!]
\centering
\includegraphics[width=0.48\textwidth]{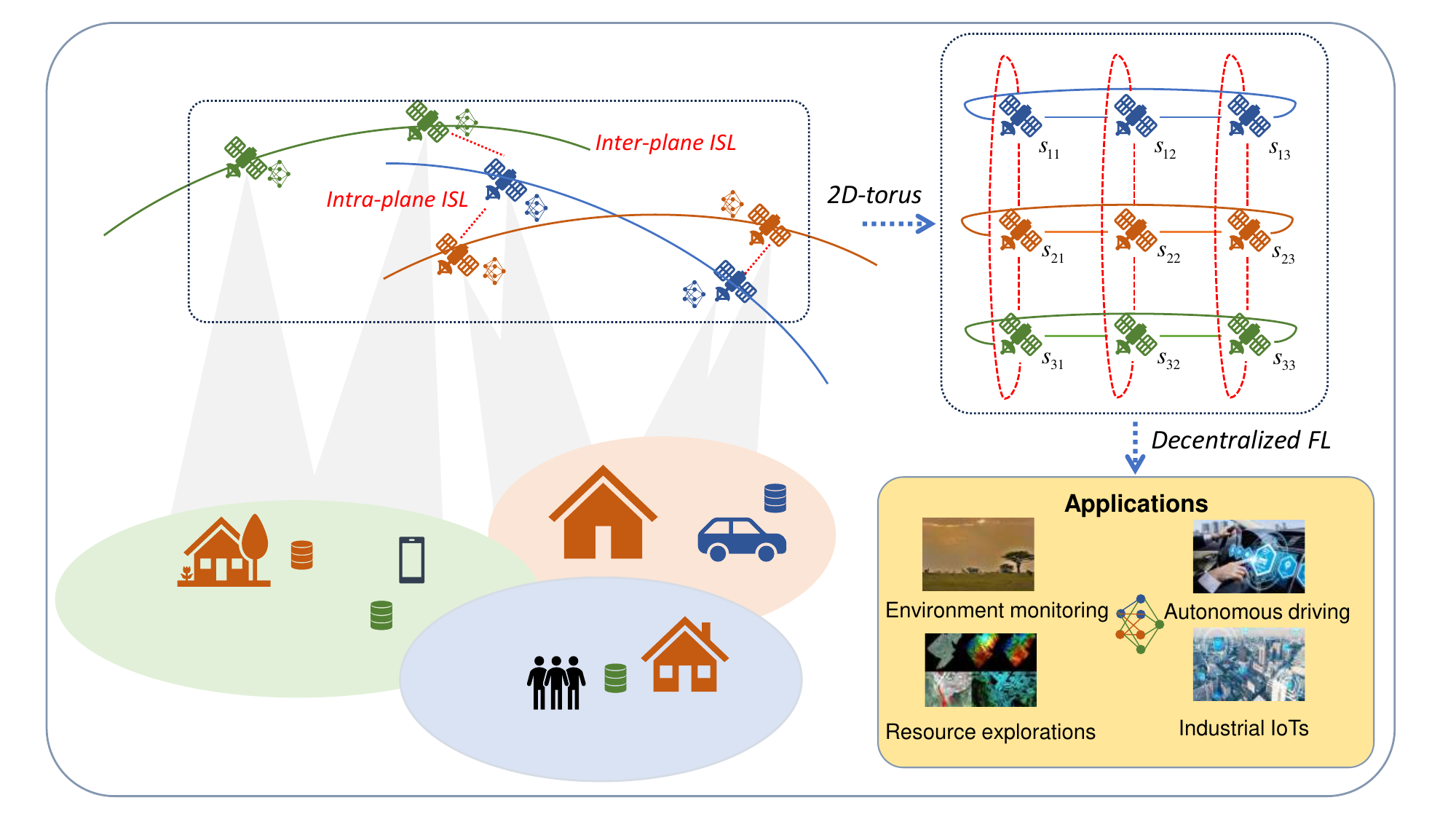}
\caption{An illustration of the LEO system.}
\label{torus}
\end{figure}

We consider a general LEO constellation consisting of $M$ planes, each with $K$ satellites. The satellites are represented by the set $\mathcal{N}=\{s_{11},...,s_{1K},...,s_{M1},..., s_{MK}\}$, where each element $s_{mk}$ denotes the $k$-th satellite in the $m$-th plane. These satellites are evenly spaced by $\pi/M$ radians within each plane, ensuring an equal number of satellites per plane. 

Each LEO satellite is equipped with four antennas, two on the pitch axis and two on the roll axis, facilitating unicast satellite-to-satellite communication. This configuration provides four ISLs: two intra-plane ISLs for communication within the same orbital plane and two inter-plane ISLs for communication with satellites in different planes. Consequently, we can model the LEO satellite network as a 2D-torus undirected graph $\mathcal{G}=(\mathcal{N},\mathcal{E})$, where we have $|\mathcal{N}| = MK$ and $\mathcal{E}$ represents the set of ISLs. An illustration of the LEO system is shown in Fig. \ref{torus}. 


\subsection{Satellite Communication Model}

Compared to radio frequency communication, free-space optical communication is more suitable for interplanetary communication scenarios due to its higher data rate and stronger anti-jamming capabilities. Therefore, we consider laser communication for ISLs between satellites. The received signal power $P_R$ can be mathematically expressed as 
\begin{equation}\label{eq1}
P_{R}=P_{T}\eta_{T}G_{T}\eta_{R}G_{R}L_{T}L_{R}\biggl(\frac{\lambda}{4\pi l}\biggr)^{2},
\end{equation}
where $P_{T}$ is the transmitting optical power, $\eta_{T}$ and $G_{T}$ are the transmitting optical efficiency and telescope gains, $\eta_{R}$ and $G_{R}$ are the receiving optical efficiency and telescope gains, and $L_{T}$ and $L_{R}$ are the transmitter and receiver pointing loss factors, respectively\cite{polishuk2004optimization}. $\lambda$ is the wavelength and $l$ is the distance between the transmitter and receiver. If the beam is Gaussian, the pointing loss factors can be expressed as
\begin{equation}  
L_{T}=e^{-G_T\theta_T^2},L_{R}=e^{-G_R\theta_R^2},
\end{equation}
where $\theta_T$ and $\theta_R$ are the transmitter and receiver radial pointing error angles, respectively.
Assuming the transmitter and receiver use telescopes with the same diameter $D$, we have
\begin{equation}  
G_{T}=G_{R}=G\approx\bigg({\frac{\pi D}{\lambda}}\bigg)^{2}.
\end{equation}
Substituting this into Eq.~(\ref{eq1}), we can obtain
\begin{equation}
P_{R}=P_{T}\eta_{T}\eta_{R}G^2e^{-G(\theta_T^2+\theta_R^2)}\biggl(\frac{\lambda}{4\pi l}\biggr)^{2},
\end{equation}
where $\theta_T^2+\theta_R^2$ follows a chi-squared distribution with four degrees of freedom, since $\theta_T$ and $\theta_R$ each encompass two normal processes (azimuth and elevation) \cite{polishuk2004optimization}. Assuming the standard deviation of the pointing error angles is equal, i.e., $\sigma_T=\sigma_R=\sigma_\theta$, the sum $\theta_T^2+\theta_R^2$ follows a gamma distribution $\Gamma\sim\{2,\frac{1}{2\sigma_\theta^2}\}$.

Furthermore, we analyze three primary forms of noise present in a laser communication system: signal shot noise, dark current noise, and thermal-Johnson noise, each with their respective variances as follows:
\begin{equation}\label{n3}
\sigma_{{sn}}^{2}=2qR_{p}P_{R}B,~~
\sigma_{{dc}}^{2}=2qI_{d}B,~~
\sigma_{th}^{2}={\frac{4k_{B}T_nB}{R_L}}, 
\end{equation}
where $q$ is the electron charge, $R_p$ is the responsivity, $B$ is  the bandwidth, $I_{d}$ is dark current, $k_{B}$ is the Boltzmann constant, $T_n$ is noise temperature and $R_{L}$ is the load resistance\cite{kahraman2021investigation}. This yields the signal-to-noise ratio (SNR) for ISL given as 
\begin{equation}
    \mathrm{SNR} = \frac{P_{R}}{\sigma_{dc}^2+\sigma_{th}^2+\sigma_{sn}^2}.
\end{equation}
Hence, the achievable data rate is given by:
\begin{equation}
    r = B\log_2(1+\mathrm{SNR}).
\end{equation}


\subsection{Satellite Computing Model}

We now present the DFL architecture within the aforementioned LEO constellation framework. In this setup, satellites function as clients, with each satellite $s_{mk}$ maintaining a locally distributed dataset ${\mathcal{D}_{mk}}$.  
The objective is to collaboratively train a global ML model $\mathbf{w}$ using local data by minimizing the global loss function $F\left( \mathbf{w} \right)$, defined as 
\begin{equation}
\mathop{\min}\limits_{\mathbf{w}\in\mathbb{R}^d}  F\left( \mathbf{w} \right) \triangleq \sum\limits_{s_{mk}\in \mathcal{N}}\frac{|\mathcal{D}_{mk}|}{|\mathcal{D}|} { {f_{mk}}\left( {\mathbf{w}} \right)},
\end{equation}
where $|\mathcal{D}|=\sum_{s_{mk}\in \mathcal{N}}{|\mathcal{D}_{mk}|}$ is the total number of train data samples across all satellites. The local loss function $f_{mk}$ is defined as:
\begin{equation}
{f_{mk}}\left( \mathbf{w} \right) = \frac{1}{{\left| {{\mathcal{D}_{mk}}} \right|}}\sum\limits_{\xi \in {\mathcal{D}_{mk}}} {{f_{mk}}} \left( \mathbf{w}_{mk},\xi  \right),
\end{equation}
where ${{f_{mk,\xi }}} \left( \mathbf{w}_{mk} \right)$ is the loss function at the data point $\xi$ for the local model $\mathbf{w}_{mk}$ of satellite $s_{mk}$.

Particularly, at each training round $t$, the following steps are executed:

1) Orbit computing: By setting $ \mathbf{w}_{mk}^{t,0}= \mathbf{w}_{mk}^{t}$, each satellite $s_{mk}$ operates $I$ local epochs. The updating rule for stochastic gradient descent (SGD) is given as
 \begin{equation}\label{local computing}
  \mathbf{w}_{mk}^{t,i+1}=\mathbf{w}_{mk}^{t,i}-\eta\nabla f(\mathbf{w}_{mk}^{t,i},\xi_{t}^i), i = 0,1,...,I-1,
\end{equation}
where $\eta$ is the learning rate and $\nabla f(\cdot)$ is the stochastic gradient of satellite $s_{mk}$ with respect to the mini-batch $\xi_{t}^i$, sampled from the local dataset ${\mathcal{D}_{mk}}$.
As a result, each satellite $s_{mk}$ obtains the updated local model $\mathbf{w}_{mk}^{t,I}$.

2) Model consensus: Each satellite $s_{mk}$ exchanges its updated local model $\mathbf{w}_{mk}^{t,I}$ with its neighbor satellites $\mathcal{N}_{mk}$ via intra-plane and inter-plane ISLs. This process completes the model consensus step, given as:
 \begin{equation}
  \mathbf{w}_{mk}^{t+1}=\sum_{s_{nj}\in \mathcal{N}_{mk}}\frac{|\mathcal{D}_{nj}|}{|\mathcal{D}_{\mathcal{N}_{mk}}|} \mathbf{w}^{t,I}_{nj},
\end{equation}
where we have $|\mathcal{D}_{\mathcal{N}_{mk}}| = \sum_{s_{nj}\in \mathcal{N}_{mk}} |\mathcal{D}_{nj}|$.

Then, the next training round $t+1$ starts. This repeats until the prespecified termination criteria are met.

In the model consensus stage,
the model exchange process can be characterized by the mixing matrix $\mathbf{Q}$. Specifically, let $\mathbf{W}^{t + 1/2}=[\mathbf{w}^{t,I}_{11},...,\mathbf{w}^{t,I}_{1K},...,\mathbf{w}^{t,I}_{M1},...,\mathbf{w}^{t,I}_{MK}]^\top \in \mathbb{R}^{MK \times d_w}$ denote the updated model parameters of all satellites $\mathcal{N}$ in round $t$, where $d_w$ is the model dimension. The aggregated model $\mathbf{W}^{t+1}=[\mathbf{w}^{t+1}_{11},...,\mathbf{w}^{t+1}_{1K},...,\mathbf{w}^{t+1}_{M1},...,\mathbf{w}^{t+1}_{MK}]^\top$ can then be expressed as
\begin{equation}\label{W}
    \mathbf{W}^{t+1}=\mathbf{Q}\mathbf{W}^{t+1/2}.
\end{equation}
We recall the definition of the mixing matrix $\mathbf{Q}$.

\textbf{Definition 1.} \textit{(Mixing matrix).} The mixing matrix $\mathbf{Q}=[q_{mk,nj}]\in \mathbb{R}^{MK\times MK}$ in the LEO satellite network graph $\mathcal{G}=(\mathcal{N},\mathcal{E})$ is assumed to have the following properties:

$\bullet$ Every entry of $\mathbf{Q}$ satisfies $q_{mk,nj}\in [0,1]$, if $mk \neq nj$ and $(mk,nj) \notin \mathcal{E}$, $q_{mk,nj}=0$, otherwise, $q_{mk,nj} = \frac{|\mathcal{D}_{nj}|}{|\mathcal{D}_{\mathcal{N}_{mk}}|} $.

$\bullet$ $\mathbf{Q1}_{MK}=\mathbf{1}_{MK}$ and $\mathbf{Q}=\mathbf{Q}^\top$ where we have the vector $\mathbf{1}_{MK}=[1,1,...,1]^\top \in \mathbb{R}^{MK}$;

$\bullet$ Denote its eigenvalues by $1=|\lambda_1|>|\lambda_2|\geq...\geq|\lambda_{MK}|\geq 0$ and then 
there exists a fixed constant $\lambda:=\textrm{max}\{|\lambda_2|,|\lambda_{MK}|\}$ such that $\|\mathbf{Q}-\frac{1}{MK}\mathbf{1}_{MK}\mathbf{1}_{MK}^\top\|_2 \leq \lambda$.

The constant $1-\lambda$ is referred to as the spectral gap of $\mathbf{Q}$, serving as a measure of the graph's connectivity. A well-connected graph is indicated by $\lambda \xrightarrow{} 0$, implying $\mathbf{Q} \xrightarrow{} \frac{1}{MK}\mathbf{1}_{MK}\mathbf{1}_{MK}^\top$, whereas a poorly connected graph corresponds to $\lambda \xrightarrow{} 1$, implying  $\mathbf{Q} \xrightarrow{} \mathbf{I}$. The value of $\lambda$ determines the convergence speed of the Markov chain introduced by the mixing matrix $\mathbf{Q}$ to a stable state, which signifies distributed consensus.
\begin{figure*}[tp]
\centering
\includegraphics[width=0.9\textwidth]{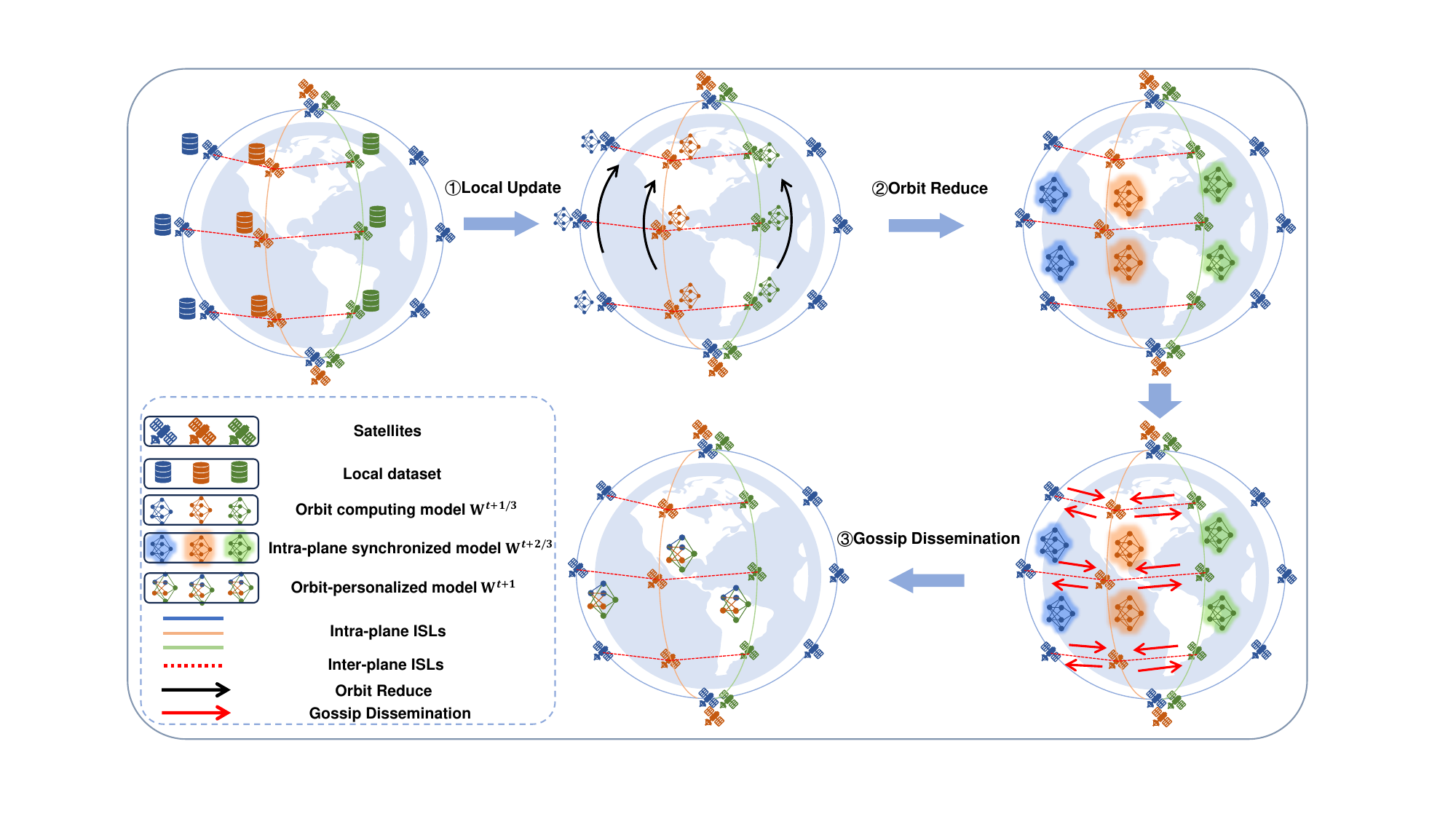}
\caption{The pipeline of DFedSat\protect\footnotemark.}
\label{pipeline}
\end{figure*}

\subsection{ISL Packet Failure Model}


The inter-plane ISLs between satellites in different planes are highly unstable due to Doppler shifts and pointing failures. To quantify the impact of this instability on model transmission, we consider a packet failure model and define the probability of successful transmission as follows:
\begin{equation}
    p := \mathbb{P}[\mathrm{SNR}>\gamma_{th}],
\end{equation}
where $\gamma_{th}$ is the link SNR threshold. To this end, each local model parameter $\mathbf{w}_{mk}^{t,I} $ is partitioned into $d$ packets and transmitted sequentially. We denote the model packets sequence as $\check{\mathbf{w}}\in \mathbb{R}^{\frac{d_w}{d} \times d}$. Following the modeling of \cite{wu2023topology}, the process of randomly receiving model packets is represented by a Bernoulli random vector $\mathbf{m} \in \mathbb{R}^{d}$, with each element $\mathbf{m}[j]\overset{i.i.d.}{\sim} \text{Bernoulli}(p), \forall j\in  [1,2,\cdots,d]$. Specifically, for any $j$, we have 
\begin{equation}
    \mathbf{m}[j]=\begin{cases} {{1, \textrm{with probability}}} & p, \\ {{0, \textrm{with probability}}} & 1-p. \\ \end{cases} 
\end{equation}
As a result, the packets received by the receiver, which have been transmitted over inter-plane ISLs, can be represented as
$\mathbf{m} \odot \mathbf{\check{w}}$.

\section{DFedSat}
In this section, we introduce our proposed algorithm, DFedSat. The key innovation is the incorporation of a two-phase aggregation strategy in the model consensus stage to address the varying stability of ISLs, thereby enhancing the efficiency of the training process. Specifically, DFedSat employs an orbit reduce mechanism for intra-plane model synchronization. In contrast, for the volatile inter-plane ISLs, DFedSat applies a flexible gossip approach to manage model dissemination across different planes. This adaptive strategy not only facilitates model diffusion but also leverages the information gleaned from the preceding intra-plane synchronization, thereby accelerating model convergence.

Additionally, to mitigate the effects of packet failures due to the instability of inter-plane ISLs, DFedSat incorporates a model self-compensation mechanism during the inter-plane model dissemination phase. This feature leverages local parameters for adaptive self-compensation, significantly enhancing the system's robustness.
\footnotetext{DFedSat is a versatile satellite FL framework applicable to various types of constellations. For clarity and brevity, we illustrate DFedSat's pipeline using the example of a specific Walker-Star constellation.}

\begin{algorithm}[pt]
\SetKwInOut{KwIn}{Input}

\renewcommand{\thealgocf}{1}
\renewcommand{\algorithmcfname}{Algorithm}
\caption{Workflow of DFedSat}\label{DFedSat}

  \KwIn{model parameters $\mathbf{w}^0$, learning rate $\eta$, local epochs $I$, gossip round $C$, total number of iterations $T$}
  $\rhd$\textrm{Model Initialization:} $\mathbf{w}_{mk}^{0,0} \xleftarrow{}\mathbf{w}^0$;
  
    \For{$t \gets 0 $  to  $T-1$}{
        \For{\rm{each satellite} $s_{mk}$ in \rm{plane} $m$ \textbf{in parallel}}{
        $\rhd$ \textrm{Orbit Computing}\textbf{ (Local Update via SGD)}\\
        \For{\rm{epoch} $i \gets 0 $  to  $I-1$}{
            
            $\mathbf{w}_{mk}^{t,i+1} = \mathbf{w}_{mk}^{t,i}-\eta *\nabla f(\mathbf{w}_{mk}^{t,i},\xi_{t}^i)$.
            }
        $\mathbf{w}_{mk}^{t+1/3}\gets \mathbf{w}_{mk}^{t,I}$.\\
    $\rhd$ \textrm{Orbit Reduce}\textbf{ (Intra-plane ISLs)}\\
    Do orbit reduce in plane $m$:\\
    $\mathbf{w}_{mk}^{t+2/3}=\sum\limits_{j=1}^{K}  q_{mk,mj}^{a}*\mathbf{w}_{mj}^{t+1/3}$.
    }
    
    
    $\rhd$ \textrm{Gossip Dissemination}\textbf{ (Inter-plane ISLs)}\\
    \For{\rm{each satellite} $s_{mk}$ in \rm{plane} $m$ \textbf{in parallel}}{
        \For{$c \gets 0 $  to  $C-1$}{
        Do model self-compensation after receiving packets from adjacent orbits via (\ref{self-comp}).\\
        Aggregate models:\\
        $\mathbf{w}_{mk}^{t+2/3,c+1}=q_{mk,mk}^{r}*\mathbf{w}_{mk}^{t+2/3,c}+q_{mk,m_lk}^{r}*\mathbf{\hat{w}}_{m_lk\xrightarrow{}mk}^{t+2/3,c}+q_{mk,m_rk}^{r}*\mathbf{\hat{w}}_{m_rk\xrightarrow{}mk}^{t+2/3,c}
        $.
        }
        $\mathbf{w}_{mk}^{t+1} \xleftarrow{} \mathbf{w}_{mk}^{t+2/3,C}$.
    
    }

}    
\end{algorithm}
\subsection{The design of DFedSat}

The DFedSat procedure consists of two phases for each round $t$: \textit{orbit computing} and \textit{model consensus}. The model consensus phase, guided by an adaptive aggregation mechanism, is further divided into two stages: intra-plane model synchronization, and inter-plane model dissemination. The pipeline of DFedSat is illustrated in Fig.~\ref{pipeline}, and the detailed steps are outlined in Algorithm \ref{DFedSat}. The specific procedures for each round $t$ in DFedSat are as follows.


\textbf{1) Orbit Computing.} Each satellite performs $I$ local epochs via (\ref{local computing}). In DFedSat, the updated model parameters serve as the starting point for the subsequent two-stage model consensus. Thus, we denote $\mathbf{w}^{t,I}_{mk}$ as $\mathbf{w}^{t+1/3}_{mk}$, and hence represent the updated parameters as $\mathbf{W}^{t+1/3}=[\mathbf{w}^{t+1/3}_{11},...,\mathbf{w}^{t+1/3}_{1K},...,\mathbf{w}^{t+1/3}_{M1},...,\mathbf{w}^{t+1/3}_{MK}]^\top$. 


\textbf{2) Model Consensus.} The entire phase encompasses intra-plane model synchronization using the ring all-reduce mechanism \cite{gibiansky2017bringing}, referred to as orbit reduce, and inter-plane model dissemination employing a flexible gossip approach. 


a) \textit{Intra-plane model synchronization}. In this stage, each plane $m$ with $K$ satellites operates independently in a ring topology. 
Our objective is for each satellite $s_{mk}$ to ultimately acquire the parameters from all other satellites in its plane $m$, thereby achieving model synchronization:
\begin{align}
\mathbf{w}_{mk}^{t+2/3}=\sum\limits_{j=1}^{K}  q_{mk,mj}^{a}*\mathbf{w}_{mj}^{t+1/3}, \forall m,k,
\end{align}
where $q_{mk,mj}^{a} = \frac{|\mathcal{D}|_{mj}}{|\mathcal{D}|_m}=\frac{|\mathcal{D}_{mj}|}{\sum_{k=1}^K|\mathcal{D}_{mk}|}$ is the intra-plane mixing weight and $|\mathcal{D}|_m$ is the total number of training data samples held by all satellites in orbit $m$.

\begin{figure}[t!]
\centering
\includegraphics[width=0.485\textwidth]{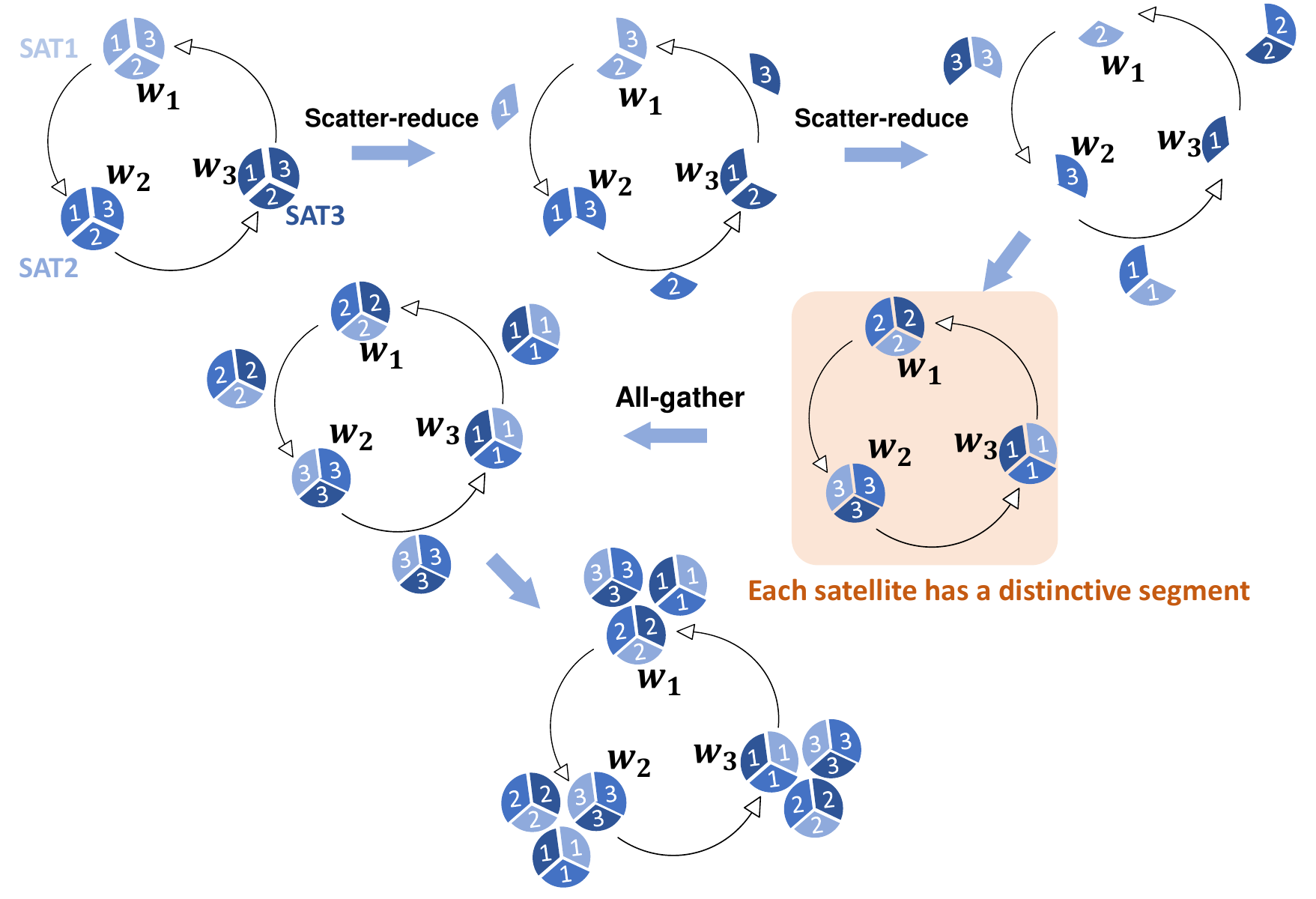}
\caption{An example of intra-plane model synchronization with $K=3$ satellites within a single orbit. $\mathbf{w}_1,\mathbf{w}_2,\mathbf{w}_3$ represent the model parameters of the three satellites, with $m$ and $t+1/3$ omitted for simplicity.}
\label{allreduce}
\end{figure}

Orbit reduce begins with the \textit{scatter-reduce} operation. Each satellite $s_{mk}$ in plane $m$ splits its local model parameter $\mathbf{w}^{t+1/3}_{mk}$ into $K$ equally-sized segments and then performs $K-1$ communication iterations. Specifically, in each iteration, each satellite $s_{mk}$ scatters one segment to the subsequent satellite while accumulating the segment received from the preceding one. After $K-1$ iterations, each satellite $s_{mk}$ holds one unique segment of $\mathbf{w}_{mk}^{t+2/3}$. Next, each satellite undergoes $K-1$ communication iterations of the \textit{all-gather} operation. During this process, each satellite replaces its corresponding segment with the received segment instead of accumulating them. At the end, each satellite $s_{mk}$ concatenates all segments to obtain the intra-plane synchronized model $\mathbf{w}_{mk}^{t+2/3}$. An example of intra-plane model synchronization for $K=3$ satellites via orbit reduce is illustrated in Fig. \ref{allreduce}.

Based on aforementioned description, the intra-plane mixing matrix $\mathbf{Q}_a = [q^{a}_{mk,nj}]\in \mathbb{R}^{MK\times MK}$ can be expressed as:
\begin{equation}
 \mathbf{Q}_a=\left( \begin{array} {c c c c } {{{\mathbf{Q}^{'}_1}}} & {{{\mathbf{0}}}} & {{{\cdots}}} & {{{\mathbf{0}}}}  \\ 
{{{\mathbf{0}}}} & {{{\mathbf{Q}^{'}_2}}} & {{{\cdots}}} & {{{\mathbf{0}}}}  \\ 
{{{\vdots}}} & {{{\vdots}}} & {{{\ddots}}} & {{{\vdots}}} \\ 
{{{\mathbf{0}}}} & {{{\mathbf{0}}}} & {{{\cdots}}} & {{{\mathbf{Q}^{'}_M}}}  \\ \end{array} \right),    
\end{equation}
where $\mathbf{Q}^{'}_{m}=[q^{a}_{mk,mj}]\in\mathbb{R}^{K \times K}$ for $1 \leq m\leq M$. $\mathbf{Q}^{'}_{m}$ represents the mixing matrix of plane $m$. Then, the intra-plane synchronized model vector $\mathbf{W}^{t+2/3}$ after orbit reduce can be express as
\begin{equation}\label{Y}
    \mathbf{W}^{t+2/3}=\mathbf{Q}_{a}\mathbf{W}^{t+1/3}.
\end{equation}

\begin{figure*}[t!]
\centering
\includegraphics[width=0.99\textwidth]{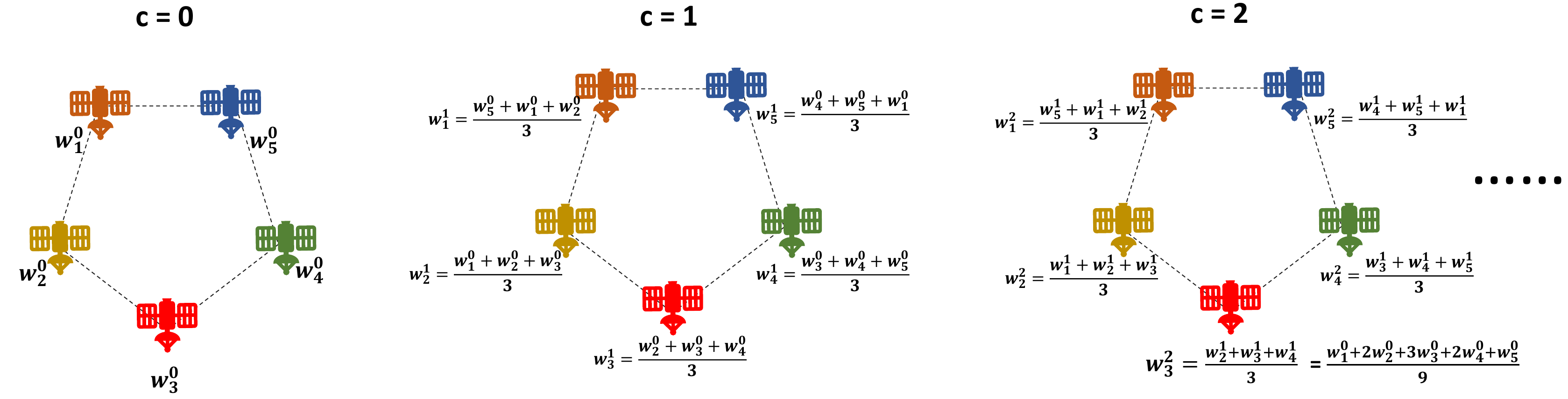}
\caption{An example of inter-plane model dissemination with $M=5$ planes. $\mathbf{w}_m^c$ represent the model parameters of the satellite in plane $m$ in the gossip round $c$, with $k$ and $t+2/3$ omitted for simplicity. Focusing on the red satellite in the plane $m = 3$, in the first gossip round $c = 1$, its model parameter $\mathbf{w}_3^1 = \frac{\mathbf{w}_2^0+\mathbf{w}_3^0+\mathbf{w}_4^0}{3}$ consists of model parameters in two other planes, and in the second gossip round $c = 2$, its model parameter $\mathbf{w}_3^2 = \frac{\mathbf{w}_2^1+\mathbf{w}_3^1+\mathbf{w}_4^1}{3} = \frac{\mathbf{w}_1^0+2\mathbf{w}_2^0+3\mathbf{w}_3^0+2\mathbf{w}_4^0+\mathbf{w}_5^0}{9}$ consists of model parameters in all other planes, which means more gossip rounds better approach the average model and enhance the model consensus.
}
\label{gossip_c}
\end{figure*}

\noindent\textbf{Remark 1.} Note that each satellite $s_{mk}$  performs a total of $2K-2$ communication iterations. During each iteration, a satellite transmits one segment, which requires only $\frac{1}{K}$ of the time needed to transmit the entire model. Consequently, the total time, given by $\frac{2K-2}{K}$, is bounded above by a constant independent of $K$. This implies that the total time required for intra-plane model synchronization remains constant regardless of the increasing scale of the constellation, demonstrating strong scalability.

b) \textit{Inter-plane model dissemination}. To adapt to the instability of the inter-plane ISLs, a flexible gossip approach is employed to aggregate the locally updated parameters from different orbits. 
Instead of fully synchronizing all satellite models within orbits as in orbit reduce, DFedSat performs partial model aggregation between orbits by gossip scheme and can flexibly adjust the extent of model diffusion by modifying the gossip round parameter $C$. Specifically, in the gossip round $c \in \{0, 1, \ldots, C-1\}
$, for satellite $s_{mk}$ in plane $m$, it receives the model parameters of the left-plane $m_l=(m-1) \% M $ and right-plane $m_r=(m+1) \% M $  satellites and aggregates them:
\begin{align}
\mathbf{w}_{mk}^{t+2/3, c + 1} &=q_{mk,mk}^{r}*\mathbf{w}_{mk}^{t+2/3,c}\nonumber\\
&+q_{mk,m_lk}^{r}*\mathbf{w}_{m_lk\xrightarrow{}mk}^{t+2/3,c}+q_{mk,m_rk}^{r}*\mathbf{w}_{m_rk\xrightarrow{}mk}^{t+2/3,c},
\end{align}
where $\mathbf{w}_{nk\xrightarrow{}mk}^{t+2/3,r}$ denotes the model parameters satellite $s_{nk}$ transmits to satellite $s_{mk}$ and $q_{mk,nk}^{r} = \frac{|\mathcal{D}_{nk}|}{|\mathcal{D}_{m_lk}|+|\mathcal{D}_{mk}|+|\mathcal{D}_{m_rk}|}$ is the inter-plane mixing weight, for $n = \{m_l, m, m_r\}$. An example of inter-plane model dissemination with $M=5$ planes is shown in Fig. \ref{gossip_c}.

After $C$ gossip rounds, each orbit has an orbit-personalized model. Similar to (\ref{Y}), the orbit-personalized model vector $\mathbf{W}^{t+1}$ can be expressed as
\begin{equation}
\mathbf{W}^{t+1}=\mathbf{Q}_{r}^{C}\mathbf{W}^{t+2/3},
\end{equation}
where $\mathbf{Q}_{r} =[q^{r}_{mk,mj}]\in\mathbb{R}^{K \times K}$ is the inter-plane mixing matrix.


\subsection{Model self-compensation for unreliable ISLs}

We proceed to consider the general scenario that the exchanged local model parameters in the inter-plane model dissemination phase could fail due to link instability.
To address this issue, DFedSat adopts a \textit{gossip self-compensation} mechanism, where
each satellite adaptively compensates for its neighbors' model parameters using corresponding parts of its local parameters.
First, satellite $s_{nk}$ partitions its intra-plane synchronized model parameters $\mathbf{w}_{nk}^{t + 2/3}$ into $d$ packets, denoted by $\mathbf{\check{w}}_{nk}^{t + 2/3}$, and transmit them to satellite $s_{mk}$.

\begin{figure}[t!]
\centering
\includegraphics[width=0.48\textwidth]{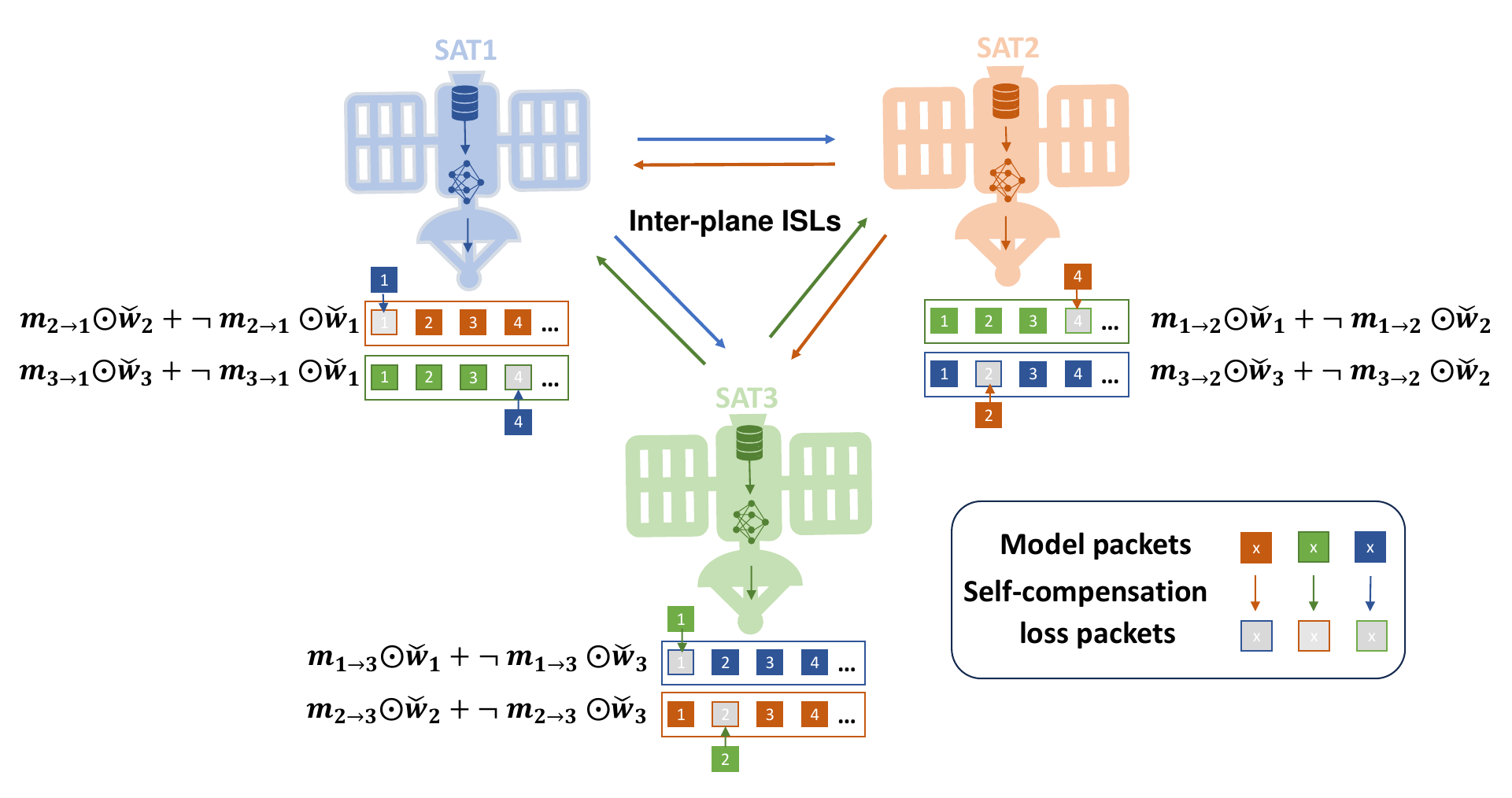}
\caption{Illustration of model gossip self-compensation scheme of satellites among $M=3$ planes. Let us focus on the process by which SAT1 receives parameter packets from SAT2 and SAT3, using the first four data packets as an example. SAT1 experiences anomalies in the first packet received from SAT2 and the fourth packet received from SAT3, where $\mathbf{m}_{2\xrightarrow{}1} = [0, 1, 1, 1]$ and $\mathbf{m}_{3\xrightarrow{}1} = [1, 1, 1, 0]$. To address these anomalies, SAT1 employs its own model's first packet $(\neg \mathbf{m}_{2\xrightarrow{}1})\odot \mathbf{\check{w}}_{1}$ and fourth packet $(\neg \mathbf{m}_{3\xrightarrow{}1})\odot \mathbf{\check{w}}_{1}$ for padding compensation. We omitted $k$ and $t+2/3$ for simplicity.}
\label{self-compensation}
\end{figure}

Then, satellite $s_{mk}$ receives the model packets $\mathbf{\check{w}}_{nk\xrightarrow{}mk}^{t+2/3}$ from satellite $s_{nk}$ by unreliable inter-plane ISL:
\begin{equation}\label{self-comp1}
    \mathbf{\check{w}}_{nk\xrightarrow{}mk}^{t+2/3} = \mathbf{m}_{nk\xrightarrow{}mk}^{t+2/3}\odot \mathbf{\check{w}}_{nk}^{t+2/3},
\end{equation}
where $\mathbf{m}_{nk\xrightarrow{}mk}^{t+2/3}$ is a Bernoulli random vector with each element adhering to the Bernoulli distribution with parameter $p_{mk,nk}$. $p_{mk,nk}$ denotes the probability of successful transmission between satellite $s_{mk}$ and $s_{nk}$. In the \textit{gossip self-compensation} scheme, the receiver $s_{mk}$ detects and sorts the received packets $\mathbf{\check{w}}_{nk\xrightarrow{}mk}^{t+2/3}$, and then fills the corrupted or lost packets with its own model $\mathbf{\check{w}}_{mk}^{t+2/3}$:
\begin{equation}\label{self-comp}
     \hat{\mathbf{w}}_{nk\xrightarrow{}mk}^{t+2/3} = \mathbf{\check{w}}_{nk\xrightarrow{}mk}^{t+2/3} + (\neg \mathbf{m}_{nk\xrightarrow{}mk}^{t+2/3})\odot \mathbf{\check{w}}_{mk}^{t+2/3}.
\end{equation}
Fig. \ref{self-compensation} is an illustration of model gossip self-compensation scheme of satellites among $M=3$ planes.

It is important to note that the compensated model $\hat{\mathbf{w}}_{nk\xrightarrow{}mk}^{t+2/3}$ becomes stochastic due to the randomness of $\mathbf{m}_{nk\xrightarrow{}mk}^{t+2/3}$. We can analyze this model diffusion process with a self-compensating mechanism from the perspective of expected values. With this rule, the expectation of orbit-personalized model vector $\mathbf{W}^{t+1}$ can be calculated as:
\begin{equation}\label{Qr}
\mathbb{E}\{\mathbf{W}^{t+1}\}=\mathbb{E}\{\mathbf{Q}_r^{C}\mathbf{W}^{t+2/3}\}=\mathbb{E}\{\mathbf{Q}_r^{C}\}\mathbf{W}^{t+2/3},
\end{equation}
where $\mathbb{E}\{\mathbf{Q}_r\}$ is defined based on (\ref{self-comp1}) and (\ref{self-comp}):
\begin{equation}
    \mathbb{E}\{\mathbf{Q}_r\}:=\left\{\begin{array}{l l}{{q^{r}_{mk,nk}p_{mk,nk},}}&{{m\not=n,}}\\ {{1-\displaystyle\sum_{k=1,n\neq m}^{M}q^{r}_{mk,nk}p_{mk,nk},}}&{{m=n.}}\end{array}\right. 
\end{equation}
From (\ref{Qr}), we can see that aggregating models via unreliable inter-plane ISLs is equivalent to reliable aggregation with the mixing matrix $\mathbb{E}\{\mathbf{Q}_r\}$. 

\textbf{Remark 2.} 
In the presence of transmission failures, instead of relying on traditional retransmission mechanisms, we utilize the corresponding information from each satellite for compensation. As illustrated in Section V, this \textit{gossip self-compensation} scheme effectively addresses packet failures, facilitating efficient convergence without significantly increasing overall communication overhead. Moreover, this approach is highly practical for satellite networks with limited resources.

\section{Convergence Analysis}
In this section, we present the convergence analysis of DFedSat for the general non-convex DFL setting. We begin by introducing the necessary assumptions.

\textbf{Assumption 1.} \textit{(L-smoothness).} For each satellite $s_{mk}\in \mathcal{N}$ and the parameter $\mathbf{w},\mathbf{v}$, the function $f_{mk}$ is differentiable and $L$-smooth, i.e., 
$$
\|\nabla f_{mk}(\mathbf{v})-\nabla f_{mk}(\mathbf{w})\|\leq L\|\mathbf{v-w}\|, \forall{\mathbf{v,w}}\in \mathbb{R}^{d_w}.
$$

The first-order Lipschitz assumption is widely used in decentralized stochastic optimization works \cite{wang2019cooperativesgdunifiedframework, 
koloskova2019decentralizedstochasticoptimizationgossip,
li2022decentralizedstochasticproximalgradient}, reflecting the functions' smoothness.

\textbf{Assumption 2.} \textit{(Uniform bounded local noise).} For each satellite $s_{mk}\in \mathcal{N}$ and all the parameter $\mathbf{w} \in \mathbb{R}^{d_w}$, the uniform local variance bound is given as
$$
\mathbb{E}\|\nabla f_{mk}(\mathbf{w},\xi)-\nabla f_{mk}(\mathbf{w})\| \leq \sigma^2.
$$

We assume uniform bounded local noise, as done in many prior works \cite{Lian_Zhang_Zhang_Hsieh_Zhang_Liu_2017},\cite{Tang_Gan_Zhang_Zhang_Ji_2018},\cite{Wang_Joshi_2018}. The uniform local variance is used for the ease of analysis and it is straightforward to generalize to non-uniform cases.

\textbf{Assumption 3.}\label{ass4.3} \textit{(Uniform bounded global noise).} For each satellite $s_{mk}\in \mathcal{N}$, the uniform global variance bound is given as
$$
\frac{1}{MK}\sum_{m=1}^{M}\sum_{k=1}^{K}\|\nabla f_{mk}({\bf w})-\nabla f({\bf w})\|\leq\zeta^{2}. 
$$

The global variance measures the degree of non-IID data distributions (the heterogeneity of data sets), as used in \cite{Reddi_Charles_Zaheer_Garrett_Rush_Konečný_Kumar_McMahan_2020}, \cite{Tian_Sahu_Zaheer_Sanjabi_Talwalkar_Smith_2018}. If the data across satellites are IID, then the upper bound of the global variance $\zeta$ is 0.


Based on the assumptions above, we present the following theorem that characterizes the convergence of DFedSat.

\textbf{Theorem 1.} Let Assumptions 1, 2, and 3 hold. 
Choosing the stepsize $\eta$ satisfies $0<\eta<\frac{1}{4LI}$, then we have
\begin{align}
        &\frac{1}{T}\sum_{t=0}^{T-1}\mathbb{E}\|\nabla f(\overline{{{\mathbf{w}^{t}}}})\|^2  \notag \\
        &\leq \frac{2f(\overline{\mathbf{w}^0})-2f(\overline{\mathbf{w}^T})}{T(\eta I-\alpha)}+\frac{\beta(\eta L^2I+L+\frac{L^2\alpha}{(1-\lambda_a\lambda_r^{C})^2})}{\eta I-\alpha},
\end{align}
where we have defined $\alpha=36\eta^3L^2I^3-36\eta^2LI^2$, $\beta=6\eta^2I^2\sigma^2+18\eta^2I^2\zeta^2$, $\lambda_a$ and $\lambda_r$ are the second largest eigenvalue of $\mathbf{Q}_{a}$ and  $\mathbb{E}\{\mathbf{Q}_r\}$ respectively.

\begin{proof}
    See Appendix D.
\end{proof}

\textbf{Corollary 1.} When we choose $\eta = \mathcal{O}(1/LI\sqrt{T})$, for a sufficiently large number $T$ of training rounds, we can derive an explicit rate from Theorem 1 as follows.
\begin{align}
        \frac{1}{T}\sum_{t=0}^{T-1 }\mathbb{E} \|\nabla f(\overline{{{\mathbf{w}^{t}}}})&\|^2 \leq \mathcal{O}\left(\frac{f(\overline{\mathbf{w}^0})-f(\overline{\mathbf{w}^T})+\zeta^2}{\sqrt{T}}\right. \notag \\
        &\left.+\frac{\sigma^2}{I\sqrt{T}}+\frac{\sigma^2+I\zeta^2}{((1-\lambda_a\lambda_r^{C})^2)IT^{3/2}}\right).
\end{align}

\textbf{Remark 3.} Our algorithm DFedSat achieves a sub-linear convergence rate in the presence of unreliable links that is equivalent to the convergence rate achieved with reliable links in recent works \cite{Sun_Li_Wang_2022, shi2023improving}. We observe that the convergence rate improves when the number $I$ of local iterations increases. With a sufficiently large $I$, the local variance diminishes and does not undermine the convergence rate, as each satellite's parameters approach a local minimizer. Moreover, the bound is dominated by $\mathcal{O}\big(\frac{\sigma^2+I\zeta^2}{(1-\lambda_a\lambda_r^{C})^2IT^{3/2})}\big)$ when $\lambda_a\lambda_r^C > 1 - \frac{1}{\sqrt{T}}$. The convergence bound becomes tighter as $\lambda_a$ and $\lambda_r$ decrease and $C$ increases, which are validated in the next section.

\section{Numerical Results}

In this section, we conduct a numerical comparison of the convergence performance of DFedSat with state-of-the-art algorithms, specifically those that adapt DSGD \cite{Lian_Zhang_Zhang_Hsieh_Zhang_Liu_2017}, DFedAvg \cite{Sun_Li_Wang_2022}, and DFedSAM \cite{Sun_Li_Wang_2022} to the scenarios under study. Additionally, we analyze the robustness of these algorithms and investigate the impact of various system parameters.

\subsection{Experimental Settings}

\textbf{LEO Constellation Configurations.} We examine a Walker-Delta constellation featuring $MK$ = 100 LEO satellites distributed across $M$ = 10 orbit planes. Satellites within each orbit are equally spaced, with each orbit positioned at an altitude of 604 km and an inclination angle of 143 degrees. The data is transmitted among satellites in 1.2 MB packets. To model packet corruption or loss, we employ randomly generated i.i.d. Bernoulli vectors. The maximum number of retransmissions for the baselines is set to 3. For readers' convenience, the pertinent parameters of the LEO constellation are summarized in Table \ref{my_tabel}.

\begin{table}[t!]
\caption{Parameters Setting}
\centering
\begin{tabular}{cc}
\toprule
\textbf{Parameters} & \textbf{Value}  \\
\midrule
Wavelength $\lambda$ & 1550 nm \\
Transmit power $P_{T}$ & 10 dBm \\
Bandwidth $B$ & 2 GHz \\
Transmitting optical efficiency $\eta_{T}$ & 0.8 \\
Receiving optical efficiency $\eta_{R}$ & 0.8 \\
Telescope diameter $D$ & 75 mm \\
Responsivity $R_{p}$ & 0.6 \\
Standard point error angle $\sigma_\theta$ & 6 $\mu$rad\\
Dark current $I_{d}$ & 1 nA \\
Noise temperature $T_n$ & 500 K \\
Load resistance $R_{L}$ & 1000 Ohm \\
ISL SNR threshold $\gamma_{th}$ & 20 dB \\
\bottomrule
\end{tabular}
\label{my_tabel}
\end{table}

\textbf{Datasets and Models.} We conduct experiments on two commonly used datasets, CIFAR-10 \cite{krizhevsky2010cifar} and CIFAR-100 \cite{krizhevsky2010cifar}, focusing on the FL task of image classification. The experiments consider both IID and non-IID data distributions. In the IID setting, the training dataset is randomly distributed across all satellites with equal local dataset sizes, and each satellite has all classes. In the non-IID setting, we employ the Dirichlet partition ${\rm Dir}(\alpha)$ with parameters $\alpha=0.6$ and $\alpha=0.3$ to represent the data heterogeneity. The backbone model used is ResNet-18, with an initial learning rate of 0.1 and a decay rate of 0.998 per communication round for all experiments. The weight decay is set to 0.001, and a mini-batch size of 64 with a momentum of 0.9 is used. The number of communication rounds is set to 300 for CIFAR-10 and 250 for CIFAR-100.

\textbf{Baselines.}
The satellite scenario presents unique challenges compared to the standard decentralized federated learning framework for terrestrial applications. Satellites, acting as clients, typically operate under energy constraints, and communication links are often unstable, particularly for inter-plane ISLs. Given the scarcity of research on decentralized satellite systems, we explore the adaptation of canonical decentralized federated learning algorithms to the satellite scenarios.

$\bullet$ DSGD: DSGD performs one local update in parallel at each satellite during the local update stage. 

$\bullet$ DFedAvg: This is the decentralized version of FedAvg. Each satellite performs multiple local updates before the communication takes place. The number $I$ of the local epoch is set to 5.

$\bullet$ DFedSAM: Apart from multiple local updates, DFedSAM applies the Sharpness Awareness Minimization (SAM) optimizer at each satellite. The hyperparameter for the perturbation radius $\rho$ of this SAM optimizer is set to 0.01, consistent with this in \cite{Sun_Li_Wang_2022}.


\subsection{Evaluation of DFedSat's Convergence Operation}

\textbf{Comparing DFedSat with the baselines.} 
We evaluate the model performance of different algorithms from three key perspectives: convergence performance, communication efficiency, and ISL robustness. To ensure a fair comparison, we set the gossip round $C$ to 1,  matching the baseline settings.

\subsubsection{Convergence Performance}










\begin{figure*}[t!]
\centering
\captionsetup[subfloat]{labelfont=normalsize,textfont=normalsize}
\subfloat[CIFAR-10]{\label{fig:a}\includegraphics[width=1\textwidth]{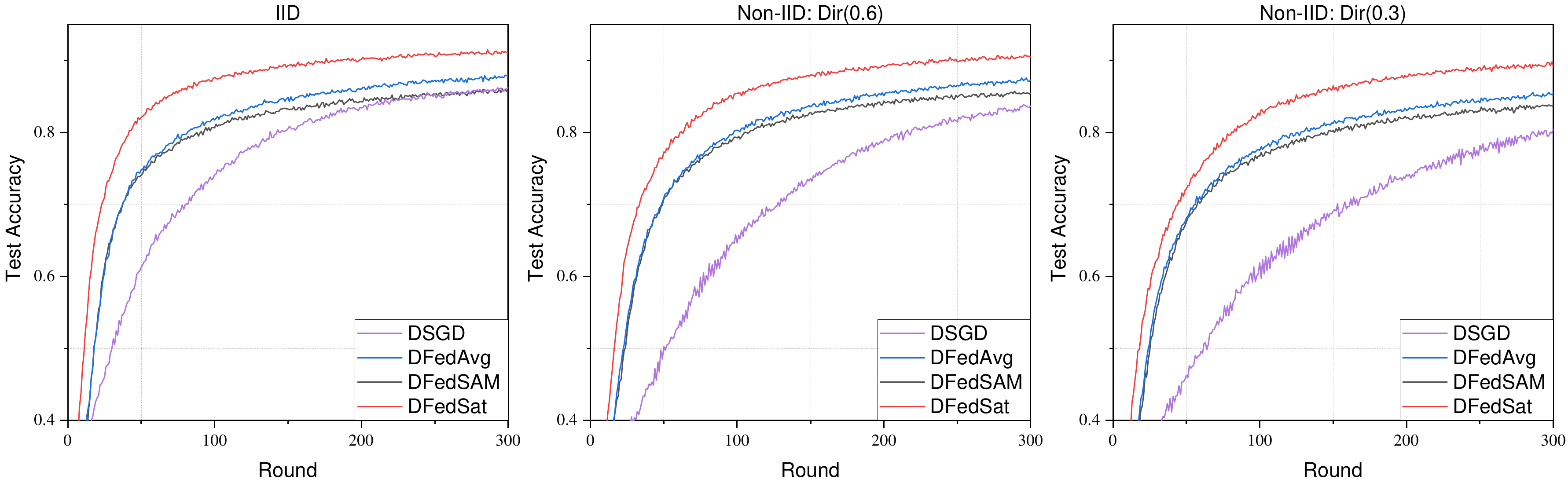}}\\
\subfloat[CIFAR-100]{\label{fig:b}\includegraphics[width=1\textwidth]{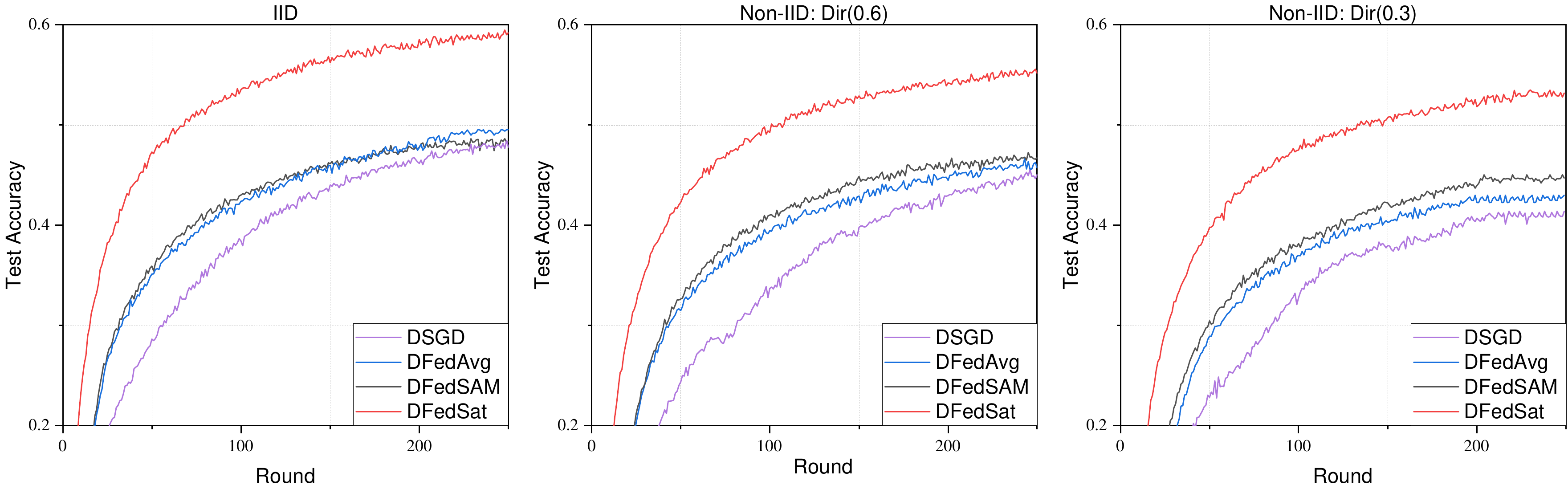}}\\	
\caption{Test accuracy for (a) CIFAR-10 and (b) CIFAR-100 under both IID and non-IID settings.}
\label{acc}
\end{figure*}

As illustrated in Fig. \ref{acc}\subref{fig:a} and \subref{fig:b}, our proposed algorithm DFedSat outperforms the baselines in terms of both test accuracy and convergence rate in both IID and non-IID settings.
The accuracy improvement is more pronounced as the level of heterogeneous data distribution increases (with $\alpha$ decreases from 0.6 to 0.3), demonstrating the effectiveness of DFedSat in addressing the statistical heterogeneity. Furthermore, DFedSat exhibits a more noticeable advantage in test accuracy on CIFAR-100 compared to CIFAR-10. This is attributed to DFedSat's two-phase model aggregation approach, where the inter-plane gossip step leverages intra-plane synchronized model information to enhance the performance of orbit-personalized models. This process effectively provides model information for non-adjacent satellites, further improving performance.


It is noteworthy that DFedAvg and DFedSAM show varying performance rankings on the two datasets under non-IID settings. Specifically, DFedAvg outperforms DFedSAM on CIFAR-10, while DFedSAM outperforms DFedAvg on CIFAR-100. This can be attributed to the SAM optimizer used in DFedSAM, which performs better with complex tasks and heterogeneous data. However, the SAM optimizer may incur more computational overhead as it requires computing the sharpness-aware gradient twice. 


\begin{figure}[t!]
\centering
\captionsetup[subfloat]{labelfont=normalsize,textfont=normalsize}
\subfloat[CIFAR-10]{\label{fig:comm-acc-a}\includegraphics[width=0.23\textwidth]{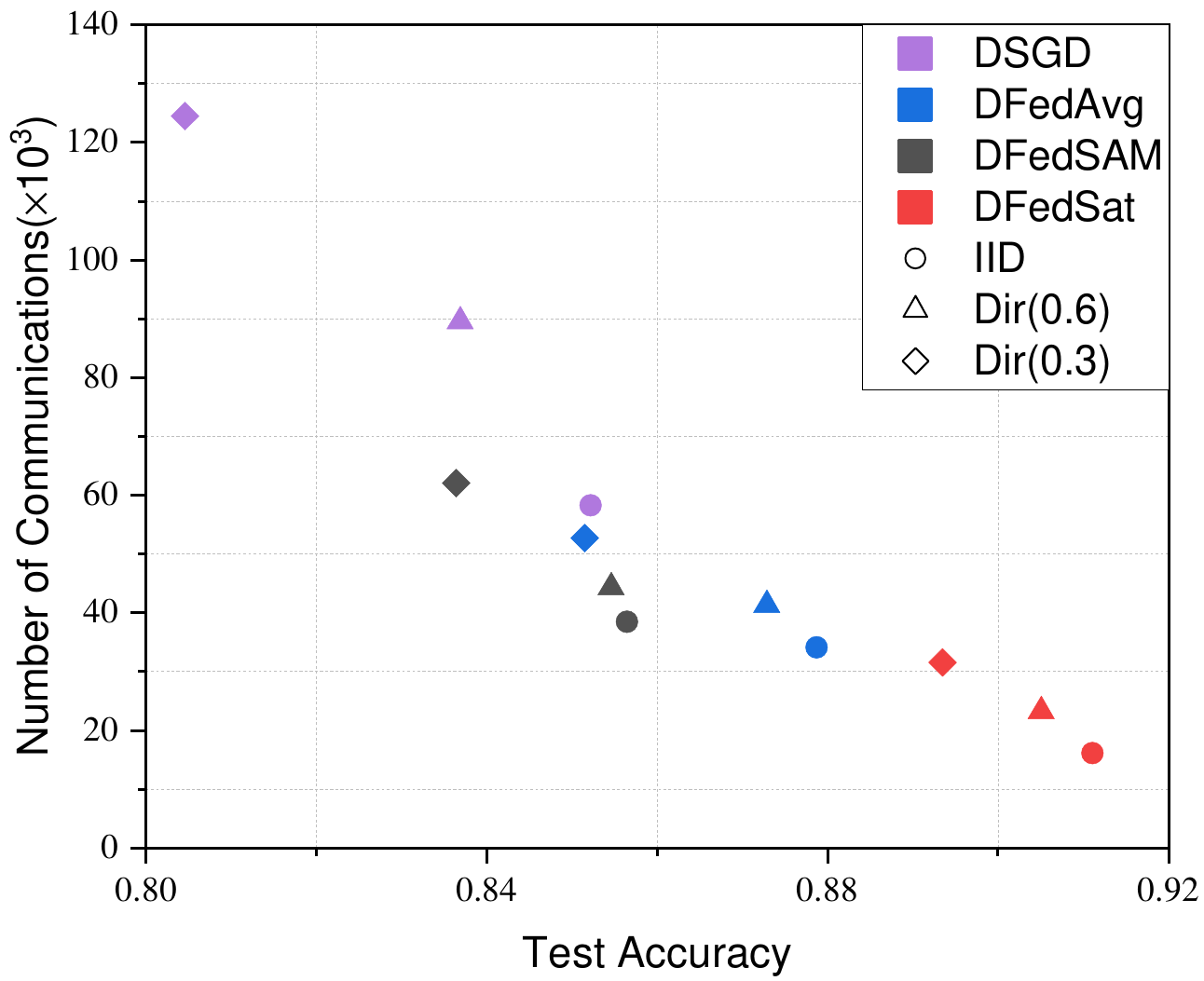}}
\subfloat[CIFAR-100]{\label{fig:comm-acc-b}\includegraphics[width=0.23\textwidth]{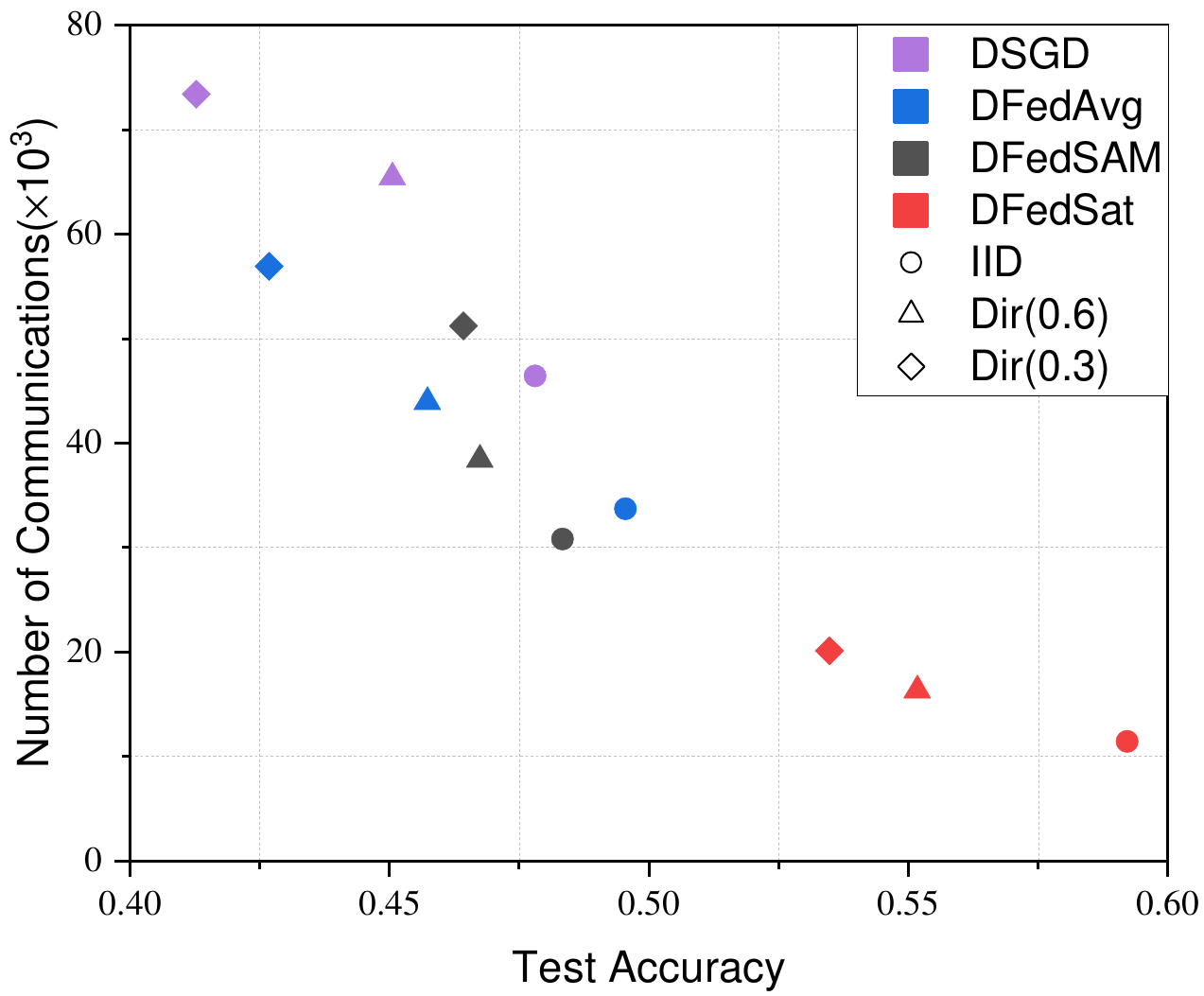}}\\	
\caption{Communication overhead for (a) CIFAR-10 and (b) CIFAR-100 under both IID and non-IID settings.}
\label{comm-acc}
\end{figure}

\subsubsection{Communication Efficiency}
Fig. \ref{comm-acc}\subref{fig:comm-acc-a} and \subref{fig:comm-acc-b} illustrate the communication overhead of all algorithms when the test accuracy reaches the target accuracy of 80\% (40\%) on CIFAR-10 (CIFAR-100) in both IID and non-IID settings. As the heterogeneous data distribution level increases (with $\alpha$ decreases from 0.6 to 0.3), the communication overhead also increases. Notably, DFedSat demonstrates a significant advantage in communication efficiency compared to the three benchmarks. Specifically, the communication overhead of DFedSat is approximately 50\% of that of DFedAvg and DFedSAM, and only 25\% of that of DSGD. This advantage can be attributed to two main factors. First, the convergence speed of DFedSat is faster than the baselines due to the two-stage aggregation, leading to a tighter upper bound of convergence and requiring fewer rounds to reach the target test accuracy. Second, the communication overhead of a single round incurred by DFedSat is lower than the baselines because the model gossip self-compensation scheme effectively mitigates the extra communication overhead of packet retransmission caused by unreliable inter-plane ISLs.

Overall, from Fig. \ref{comm-acc}\subref{fig:comm-acc-a} and \subref{fig:comm-acc-b}, we can observe the superiority of the proposed model aggregation strategy over conventional strategies. Therefore, DFedSat emerges as a more communication-efficient algorithm while maintaining high accuracy.

\begin{figure}[t!]
\centering
\captionsetup[subfloat]{labelfont=normalsize,textfont=normalsize}
\subfloat[CIFAR-10]{\label{fig:robust-a}\includegraphics[width=0.23\textwidth]{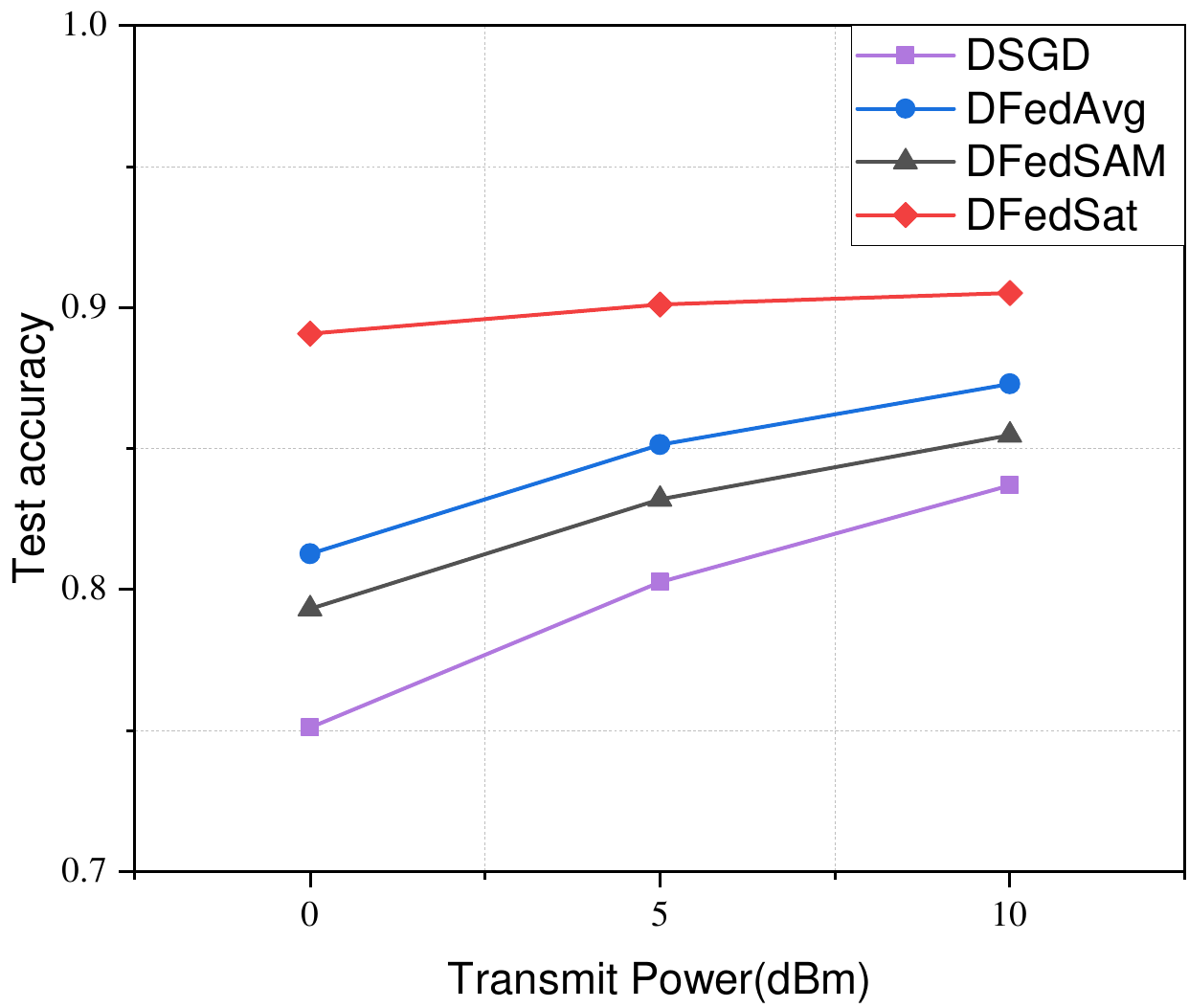}}
\subfloat[CIFAR-100]{\label{fig:robust-b}\includegraphics[width=0.23\textwidth]{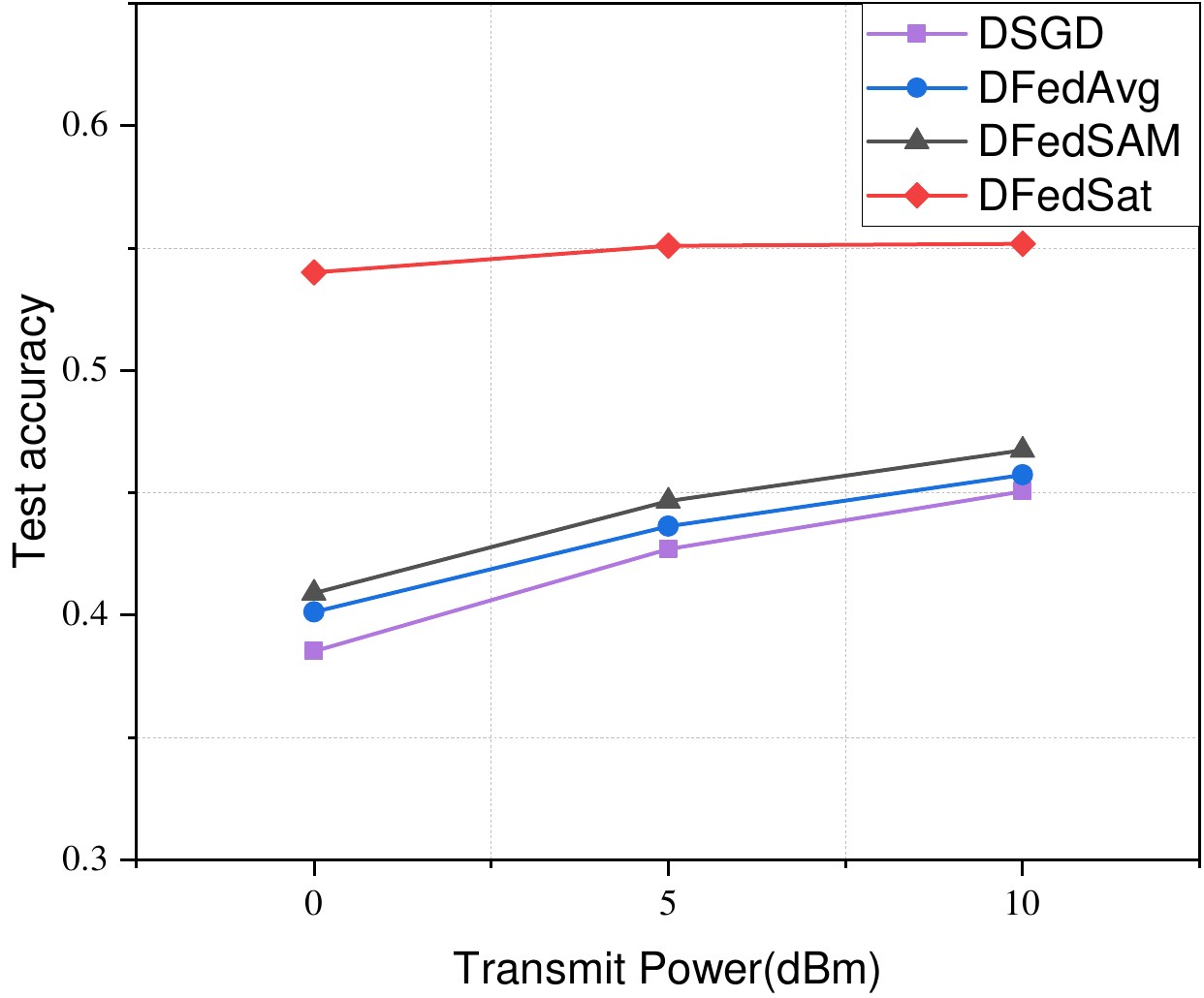}}\\	
\subfloat[CIFAR-10]{\label{fig:robust-c}\includegraphics[width=0.23\textwidth]{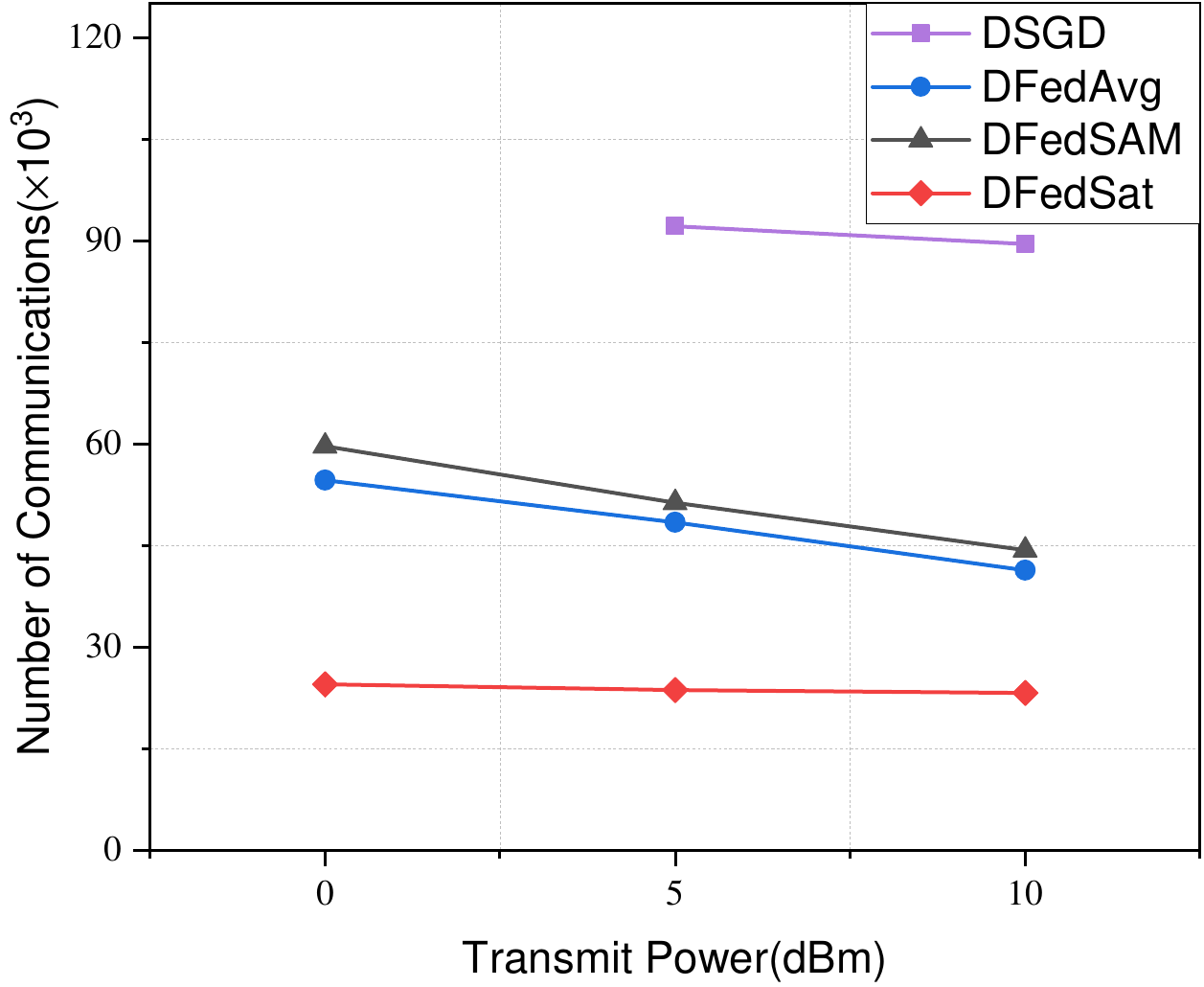}}
\subfloat[CIFAR-100]{\label{fig:robust-d}\includegraphics[width=0.23\textwidth]{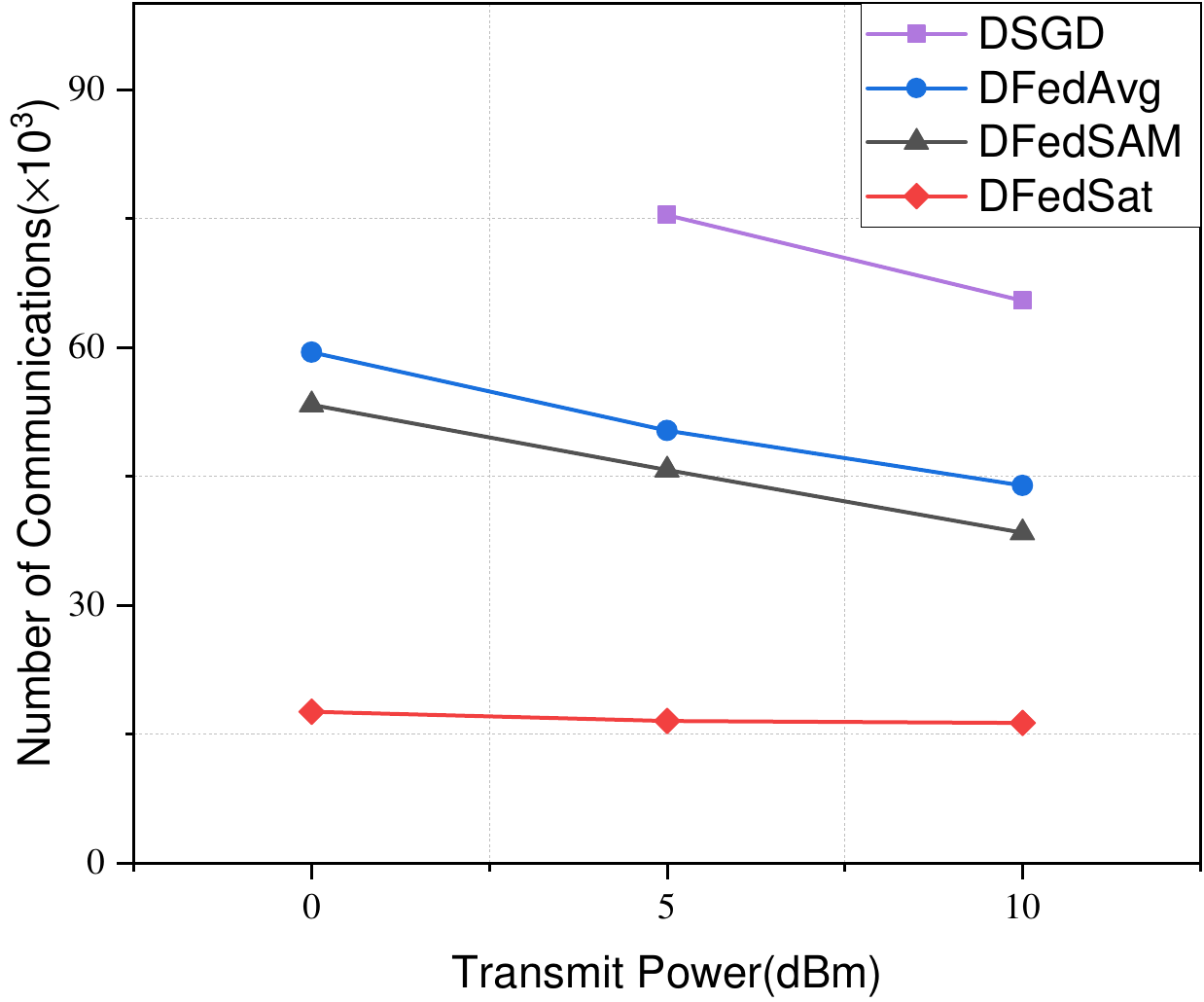}}\\
\caption{The impact of transmit power on test accuracy and communication overhead for CIFAR-10 and CIFAR-100 datasets under the non-IID setting ($\rm{Dir}(0.6)$).}
\label{robust}
\end{figure}

\subsubsection{Unreliable ISL Robustness}

Fig. \ref{robust}\subref{fig:robust-a} and \subref{fig:robust-b} show the test accuracy of each algorithm as transmit power ranging from 0 dBm to 10 dBm. The experimental results show a negligible effect on the performance of DFedSat, whereas the baselines exhibit varying degrees of decrease in test accuracy as the transmit power decreases (implying a weakening of the ISL link stability), thus highlighting the superior robustness of DFedSat. 

Fig. \ref{robust}\subref{fig:robust-c} and \subref{fig:robust-d} show the communication overhead of each algorithm under different transmit power. It should be noted that the communication overhead of DSGD in 0 dBm was not recorded because it failed to reach the target accuracy in the case of poor link quality. Similar to test accuracy, the stability of the ISL link has a negligible effect on the communication overhead because of DFedSat's model gossip \textit{self-compensation} mechanism. In contrast, the communication overhead of other baselines keeps increasing with decreasing link stability because of the repeated retransmission of erroneous or lost packets.

\begin{figure}[htbp]
\centering
\captionsetup[subfloat]{labelfont=normalsize,textfont=normalsize}
\subfloat[$\rm{Dir}(0.6)$]{\label{fig:r-a}\includegraphics[width=0.23\textwidth]{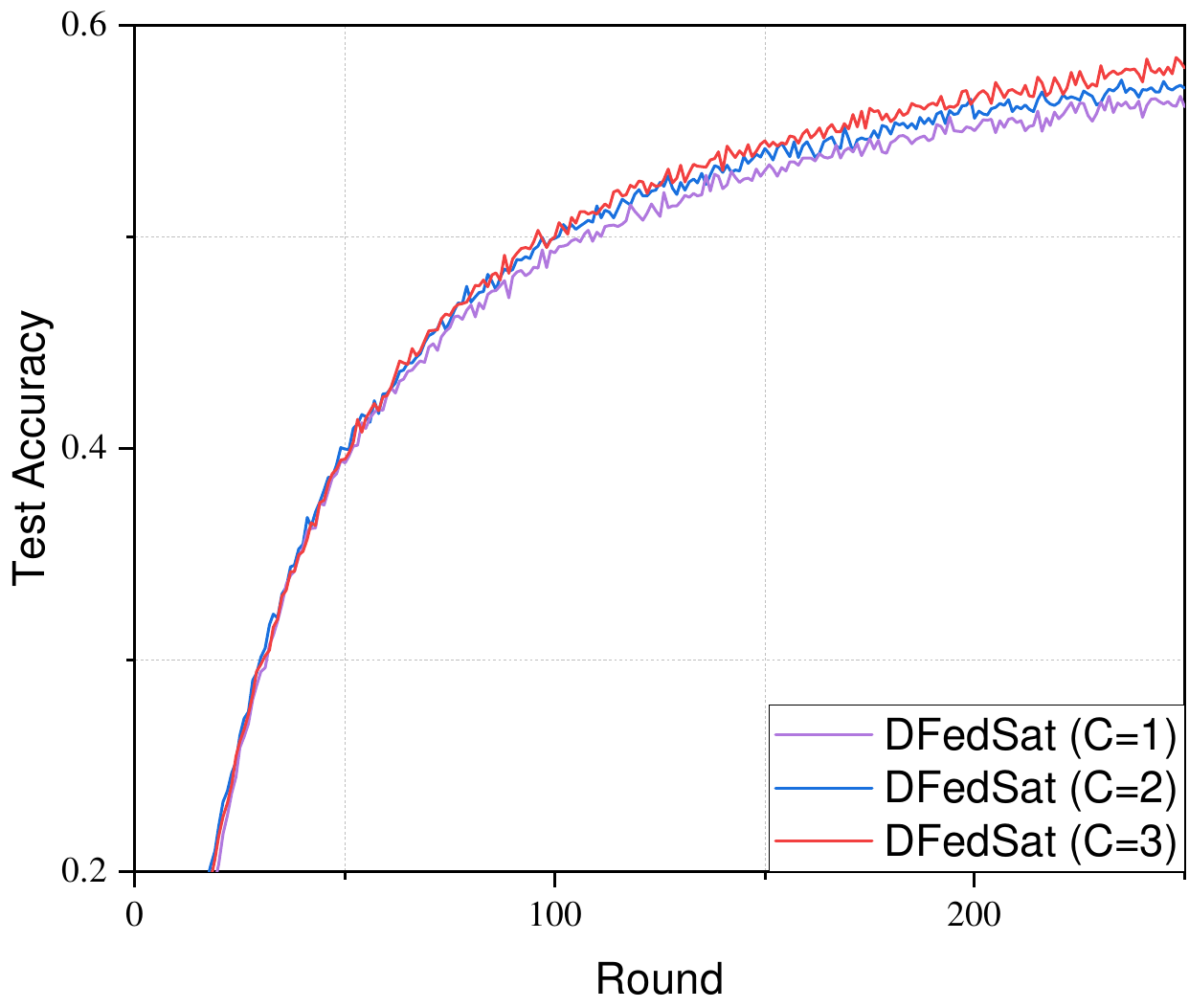}}
\subfloat[$\rm{Dir}(0.3)$]{\label{fig:r-b}\includegraphics[width=0.23\textwidth]{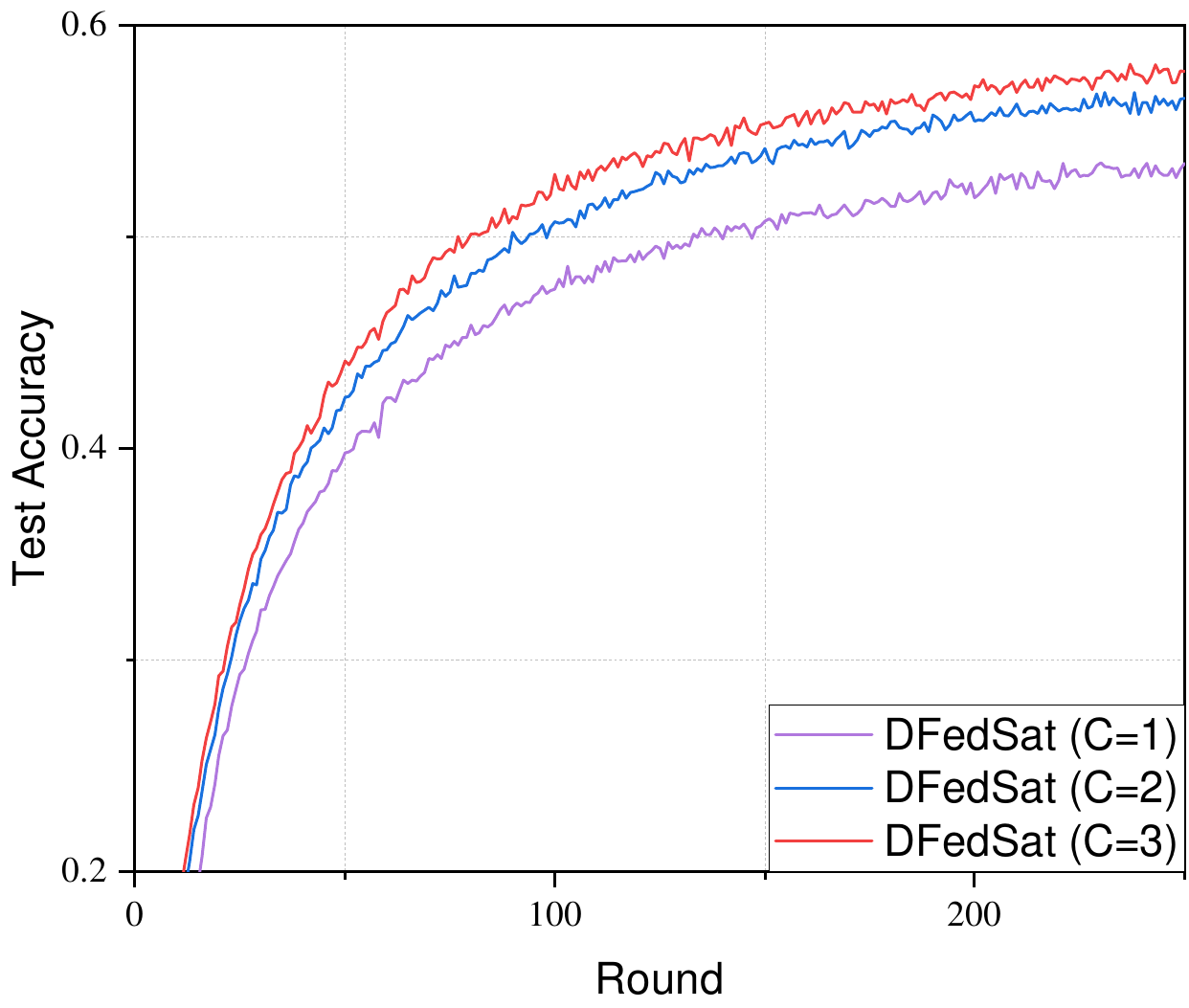}}\\	
\caption{The impact of gossip steps $C$ on test accuracy for CIFAR-100 dataset under the non-IID setting.}
\label{r-acc}
\end{figure}
\setcounter{subsubsection}{0}

\textbf{Impact of hype-parameters on DFedSat}. We conduct an in-depth analysis of how the parameters such as the gossip round $C$ and the number of planes $M$ in DFedSat influence convergence performance and communication efficiency.

\subsubsection{The impact of gossip round $C$}
Fig. \ref{r-acc}\subref{fig:r-a} and \subref{fig:r-b} illustrate the impact of gossip round $C$ on test accuracy for the CIFAR-100 dataset under the non-IID setting ($\alpha=0.6$ and $\alpha=0.3$). It is observed that larger values of $C$ can enhance performance, consistent with our theoretical findings. However, this may also lead to increased communication costs. A comparison between Fig. \ref{r-acc}\subref{fig:r-a} and Fig. \ref{r-acc}\subref{fig:r-b} reveals that the effect of increasing $C$ is more pronounced in scenarios with higher data heterogeneity. This implies that employing a larger $C$ value could serve as a beneficial strategy to effectively address the local consistency issue while striking a balance between performance and communication costs.

\begin{figure}[t!]
\centering
\captionsetup[subfloat]{labelfont=normalsize,textfont=normalsize}
\subfloat[]{\label{fig:M-a}\includegraphics[width=0.23\textwidth]{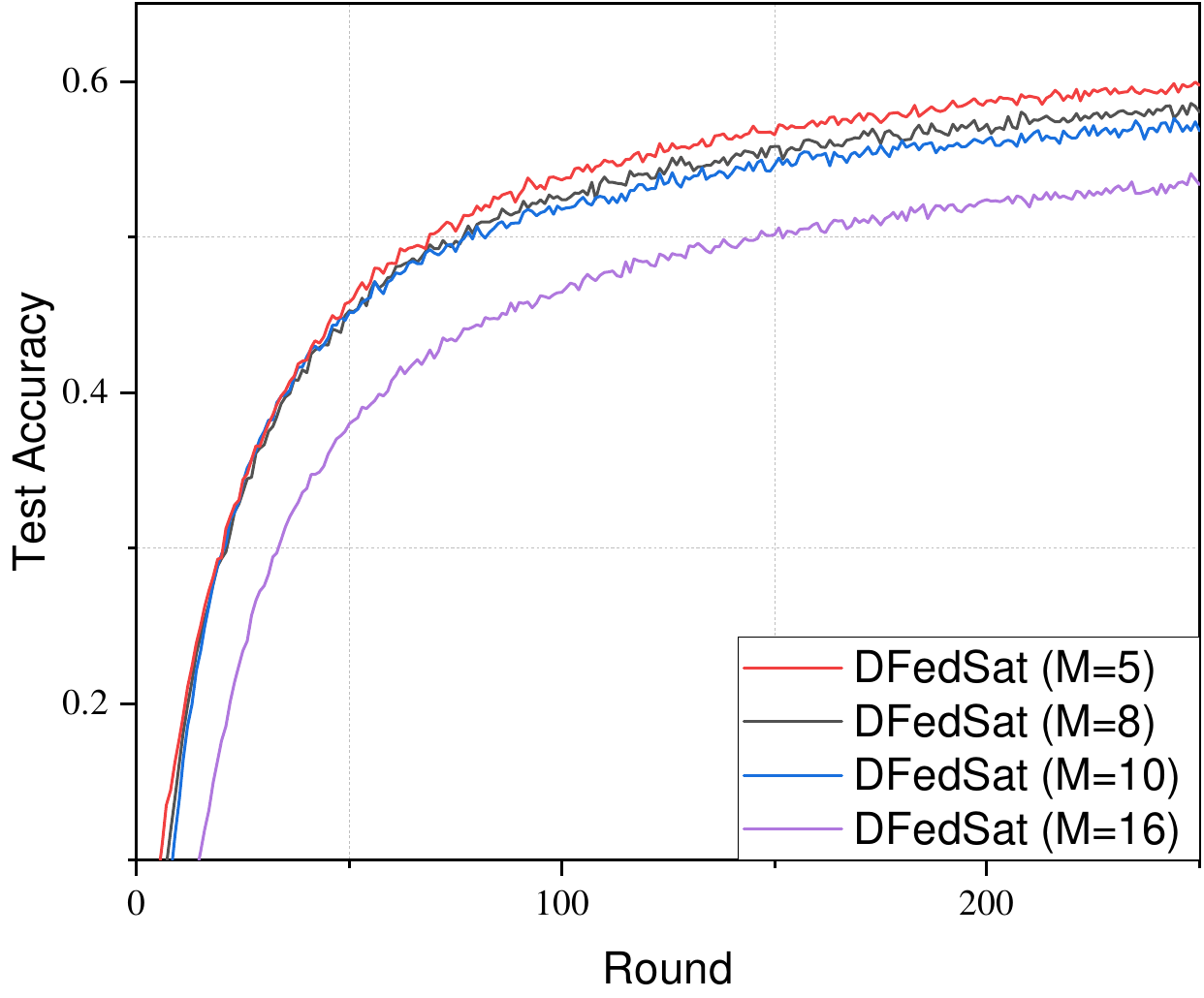}}
\subfloat[]{\label{fig:M-b}\includegraphics[width=0.23\textwidth]{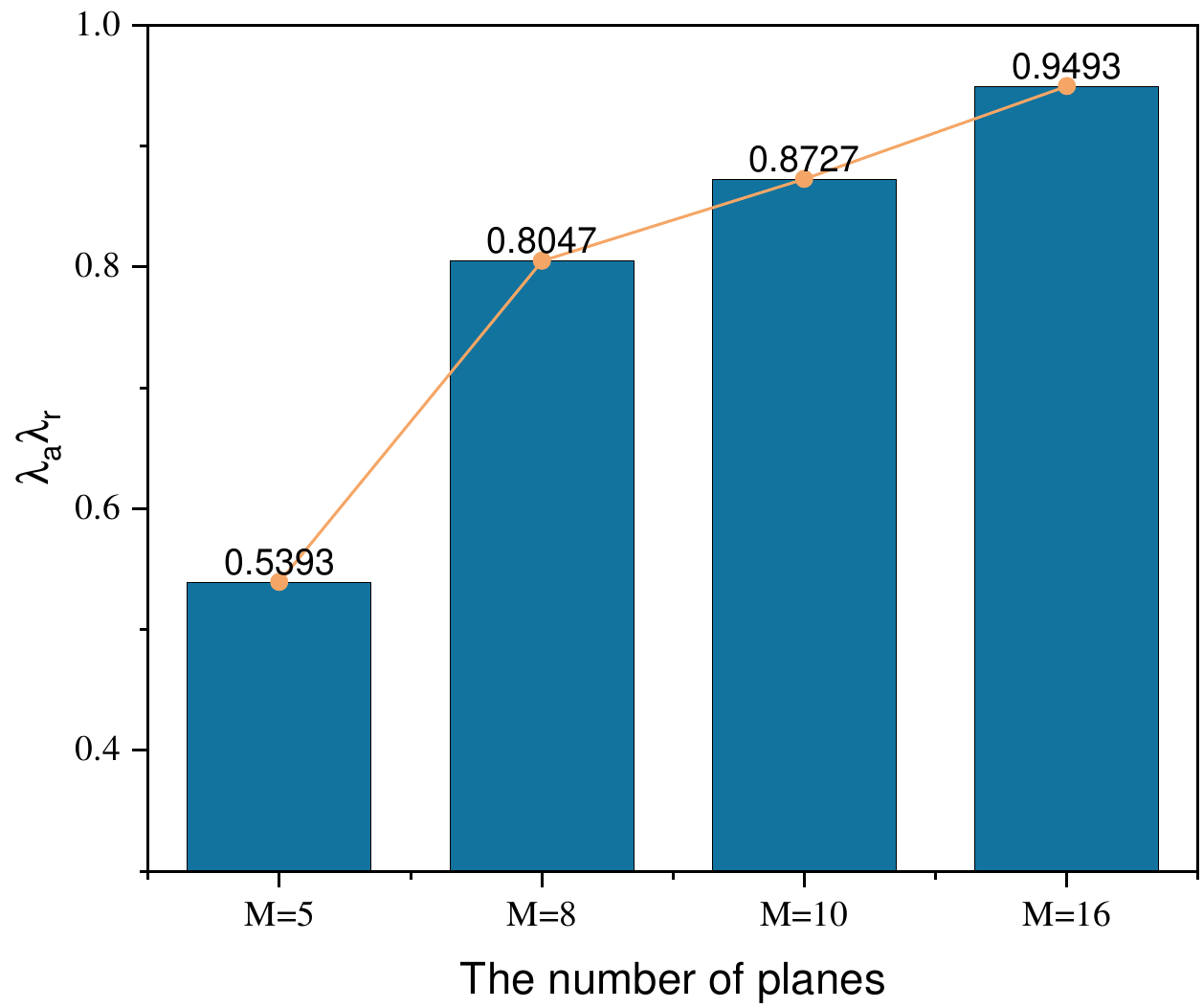}}\\	
\caption{(a): The impact of the number of planes $M$ on test accuracy for CIFAR-100 under the non-IID setting ($\rm{Dir}(0.3)$). (b): The value of $\lambda_a\lambda_r$ with different $M$}
\label{M}
\end{figure}

\subsubsection{The impact of constellation structure on $M$}
We maintain a total of 80 satellites and keep the other constellation parameters constant. As shown in Fig. \ref{M}\subref{fig:M-a}, the test accuracy tends to increase as the number of planes $M$ decreases (resulting in a higher number of satellites per plane $N$). Intuitively, with fewer planes and a higher density of satellites per plane, each satellite obtains a more comprehensive partially aggregated model. This facilitates broader dissemination of model parameters among satellites in various planes, thereby enhancing overall performance. Remarkably, in a single-plane constellation, DFedSat obviates the need for the gossip dissemination step entirely. From a theoretical standpoint, as illustrated in Fig. \ref{M}\subref{fig:M-b}, a smaller number of planes $M$ corresponds to a smaller value of $\lambda_a\lambda_r$, resulting in a tighter upper bound of convergence. Our experimental findings align with this theoretical perspective.

\section{Conclusion}
This paper has proposed DFedSat, a novel decentralized federated learning framework tailored to the LEO constellation. DFedSat has implemented two distinct mechanisms, orbit reduce and gossip dissemination, for intra-plane and inter-plane model exchanges, achieving efficient partial model aggregation. To address unreliable inter-plane model transmission, we have integrated the self-compensation mechanism into DFedSat, reducing communication costs and improving system robustness. Additionally, we have demonstrated the sublinear convergence rate for DFedSat in non-convex scenarios. Extensive experimental results have demonstrated that our algorithm performs competitively in terms of convergence performance, communication efficiency, and robustness to unreliable links, outperforming other DFL baselines

\appendices

\section{Technical Preliminaries}
\textbf{Lemma 1.} (Lemma 5, \cite{Lian_Zhang_Zhang_Hsieh_Zhang_Liu_2017}).
Define the vector $\mathbf{1}:=[1,1,...,1]^\top \in \mathbb{R}^n$ and $\mathbf{J}:=\frac{\mathbf{11}^\top}{n}\in \mathbb{R}^{n\times n}$, the mixing matrix $\mathbf{Q}\in \mathbb{R}^{n\times n}$ satisfies
\begin{equation}
\nonumber
    \| \mathbf{Q}^t-\mathbf{J}\|_{\textrm{op}}\leq \lambda^t,
\end{equation}
where $\|\cdot\|_{\textrm{op}}$ denotes the spectral norm of a matrix and $\lambda:=\textrm{max}\{|\lambda_2|,|\lambda_n(\mathbf{Q})|\}$.

\textbf{Lemma 2.} Given the stepsize $0<\eta \leq \frac{1}{4LI}$, assuming $\mathbf{w}_{mk}^{t,i}$ are generated by DFedSat for all $mk\in\{11,...,1K,...,M1,...,MK\}$ and $ 0\leq i\leq I-1$. With Assumptions 1, 2, and 3, it follows that
\begin{equation}
    \begin{split}
        &\frac{1}{MK}\sum_{m=1}^{M}\sum_{k=1}^{N}\mathbb{E}\|\mathbf{w}_{mk}^{t,i}-\mathbf{w}^t_{mk}\|^2
        \\&\leq C_1\eta^2+18I^2\eta^2\frac{\sum_{m=1}^{M}\sum_{k=1}^{N}\mathbb{E}\| \nabla f(\mathbf{w}_{mk}^{t})\|^2}{MK},
    \end{split}
\end{equation}
where $C_1=6I\sigma^2+18I^2\zeta^2$.

\textbf{Lemma 3.} Given the stepsize $\eta>0$, assuming $\mathbf{w}_{mk}^{t}$ is generated by DFedSat for all $mk\in\{11,...,1K,...,M1,...,MK\}$. With Assumptions 1, 2, and 3, it follows that
\begin{equation}
    \begin{split}
        \frac{1}{MK}\sum_{m=1}^{M}\sum_{k=1}^{K}\mathbb{E}\|\mathbf{w}^t_{mk}-\overline{\mathbf{w}^t}\|^2
        \leq C_2\frac{\eta^2}{(1-(\lambda_a\lambda_r^{C}))^2},
    \end{split}
\end{equation}
where $C_2=6I\eta^2\sigma^2+18I^2\eta^2\zeta^2+18I^2\eta^2\mathbb{E}\| \nabla f(\mathbf{w}_{mk}^{t})\|^2$.

\section{Proof of Lemma 2}
\textit{Proof:}
Note that for each satellite $s_{mk}$, for any $i\in\{0,1,...,I-1\}$, it holds
\begin{align}\label{eq28}
\mathbb{E}\|\mathbf{w}_{mk}^{t,i+1}-\mathbf{w}^t_{mk}\|^2
&=\mathbb{E}\|\mathbf{w}_{mk}^{t,i}-\eta \nabla f_{mk}(\mathbf{w}_{mk}^{t,i},\xi_{i})-\mathbf{w}^t_{mk}\|^2 \notag \\ 
&\!\!\!\!\!\!\!\!\!\!\!=\mathbb{E}\|\mathbf{w}_{mk}^{t,i}-\mathbf{w}^t_{mk} -\eta (\Lambda+\Phi+\Omega) \|^2.
\end{align}
where we have defined $\Lambda=\nabla f_{mk}(\mathbf{w}_{mk}^{t,i},\xi_{i})-\nabla f_{mk}(\mathbf{w}_{mk}^{t}), \Phi=\nabla f_{mk}(\mathbf{w}_{mk}^{t,i})-\nabla f_{mk}(\mathbf{w}_{mk}^{t}), \Omega=\nabla f_{mk}(\mathbf{w}_{mk}^{t})-\nabla f(\mathbf{w}_{mk}^{t})+\nabla f(\mathbf{w}_{mk}^{t})$.

Applying Cauchy inequality $ \mathbb{E}\|\mathbf{a}+\mathbf{b}\|^2 \leq (1+ \frac{1}{\phi})\mathbb{E}\|\mathbf{a}\|^2+(1+ \phi)\mathbb{E}\|\mathbf{b}\|^2$ to (\ref{eq28}), we can have
\begin{align}\label{eq29}
&\mathbb{E}\|\mathbf{w}_{mk}^{t,i+1}-\mathbf{w}^t_{mk}\|^2\nonumber\\
&\leq (1+\frac{1}{\phi})\mathbb{E}\|\mathbf{w}_{mk}^{t,i}-\mathbf{w}^t_{mk}-\eta\Lambda\|^2 +(1+\phi)\eta^2\mathbb{E}\|\Phi+\Omega\|^2\nonumber\\
&\overset{(a)}{\leq} (1+\frac{1}{2I-1})\mathbb{E}\|\mathbf{w}_{mk}^{t,i}-\mathbf{w}^t_{mk}\|^2+2\eta^2\sigma^2 \nonumber\\
&\quad+6I\eta^2\mathbb{E}\|\nabla f_{mk}(\mathbf{w}_{mk}^{t,i})-\nabla f_{mk}(\mathbf{w}_{mk}^{t})\|^2\nonumber\\
&\quad+6I\eta^2\mathbb{E}\|\nabla f_{mk}(\mathbf{w}_{mk}^{t})-\nabla f(\mathbf{w}_{mk}^{t})\|^2\nonumber\\
&\quad+6I\eta^2\mathbb{E}\| \nabla f(\mathbf{w}_{mk}^{t})\|^2\nonumber\\
&\overset{(b)}{\leq} (1+\frac{1}{2I-1}+6I\eta^2L^2)\mathbb{E}\|\mathbf{w}_{mk}^{t,i}-\mathbf{w}^t_{mk}\|^2\nonumber\\ &\quad+2\eta^2\sigma^2+6I\eta^2\zeta^2+6I\eta^2\mathbb{E}\| \nabla f(\mathbf{w}_{mk}^{t})\|^2\nonumber\\
&\overset{(c)}{\leq} (1+\frac{1}{I-1})\mathbb{E}\|\mathbf{w}_{mk}^{t,i}-\mathbf{w}^t_{mk}\|^2 \nonumber\\
&\quad+2\eta^2\sigma^2+6I\eta^2\zeta^2+6I\eta^2\mathbb{E}\| \nabla f(\mathbf{w}_{mk}^{t})\|^2,
\end{align}
where the inequality $(a)$ holds because of Assumption 2 and let $\phi = 2 I - 1$, $(b)$ uses the Assumption 1 and 3, and $(c)$ needs the condition that $\eta^2\leq \frac{1}{6L^2(I-1)(2I-1)}$.



Thus, based on (\ref{eq29}), the recursion from $j=0$ to $i$ yields
\begin{align}\label{30}
&\mathbb{E}\|\mathbf{w}_{mk}^{t,i}-\mathbf{w}^t_{mk}\|^2\nonumber\\
        &\leq \sum_{j=0}^{I-1}(1+\frac{1}{I-1})^j(2\eta^2\sigma^2+6I\eta^2\zeta^2+6I\eta^2\mathbb{E}\| \nabla f(\mathbf{w}_{mk}^{t})\|^2)\nonumber\\
        &\overset{(a)}{\leq} \Upsilon(I)(2\eta^2\sigma^2+6I\eta^2\zeta^2+6I\eta^2\mathbb{E}\| \nabla f(\mathbf{w}_{mk}^{t})\|^2)\nonumber\\
        &\overset{(b)}{\leq} 6I\eta^2\sigma^2+18I^2\eta^2\zeta^2+18I^2\eta^2\mathbb{E}\| \nabla f(\mathbf{w}_{mk}^{t})\|^2,
\end{align}
where $\Upsilon(I)=(I-1)((1+\frac{1}{I-1})^I-1)$, $(a)$ uses the formula for the sum of an equipartite series and $(b)$ uses the fact of $\Upsilon\leq3$ for any $I>1$. Summing $m$ from $1$ to $M$ and $k$ from $1$ to $K$ in turn and multiplying $\frac{1}{MK} $ on both sides of (\ref{30}), we can get Lemma 2.
\section{Proof of Lemma 3}


\textit{Proof:} For simplicity, we denote that 
$\mathbf{Y}^t:=\mathbf{W}^{t + 1/3} \in \mathbb{R}^{MK \times d_w}$, $\mathbb{E}\{\mathbf{Q}_r^C\}:=\overline{\mathbf{Q}_{r}^C}$. With this notion, we can have
\begin{align}\label{123}
&\mathbb{E}\{\mathbf{W}^{t+1}\}=\overline{\mathbf{Q}_{r}^C}\mathbf{W}^{t+2/3}\nonumber\\
&=\overline{\mathbf{Q}_{r}^C}\mathbf{Q}_{a}\mathbf{Y}^{t}=\overline{\mathbf{Q}_{r}^C}\mathbf{Q}_{a}\mathbf{W}^t-\zeta^t,   
\end{align}
where we have $\zeta^t:= \overline{\mathbf{Q}_{r}^C}\mathbf{Q}_{a}\mathbf{W}^t-\overline{\mathbf{Q}_{r}^C}\mathbf{Q}_{a}\mathbf{Y}^t$. 

The iteration equation (\ref{123}) can be rewritten as the following expression
\begin{equation}\label{3} 
\mathbb{E}\{\mathbf{W}^{t}\}=(\overline{\mathbf{Q}_{r}^C}\mathbf{Q}_{a})^{t}{\mathbf{W}}^{0}-\sum_{j=0}^{t-1}(\overline{\mathbf{Q}_{r}^C}\mathbf{Q}_{a})^{(t-1-j)}\zeta^{j}. 
\end{equation}

Since $\mathbf{Q}_{r}$ and $\mathbf{Q}_{a}$ are doubly stochastic matrix, then it follows
\begin{equation}\label{4}
\overline{\mathbf{Q}_{r}^C}\mathbf{Q}_{a}\mathbf{J}=\mathbf{J}\overline{\mathbf{Q}_{r}^C}\mathbf{Q}_{a}=\mathbf{J}.
\end{equation}
According to Lemma 1, it holds that
\begin{equation}\label{2}
    \|\mathbf{Q}^{t}-\mathbf{J}\|\leq \lambda^{t}.
\end{equation}
Following (\ref{2}), it holds that
\begin{equation}\label{55}
   \|(\overline{\mathbf{Q}_{r}^C}\mathbf{Q}_{a})^{t}-\mathbf{J}\|\leq (\lambda_a\lambda_r^{C})^{t},
\end{equation}
where $\lambda_a$ and $\lambda_r$ are the second largest eigenvalue of $\mathbf{Q}_{a}$ and $\overline{\mathbf{Q}_{r}}$ respectively.

Multiplying both sides of (\ref{3}) with $\mathbf{J}$ and using initialization $\mathbf{W}^0=0$, we can get that
\begin{equation}\label{44}
\mathbf{J}\mathbf{W}^{t}=\mathbf{J}\mathbf{W}^{0}-\sum_{j=0}^{l-1}\mathbf{J}\zeta^{j}=-\sum_{j=0}^{l-1}\mathbf{J}\zeta^{j}. 
\end{equation}

Then, we have
\begin{align}
   \mathbb{E}\|\mathbf{W}^{t}-\mathbf{J}\mathbf{W}^{t}\| &\overset{(a)}{=}\|\sum_{j=0}^{t-1}(\mathbf{J}-(\overline{\mathbf{Q}_{r}^C}\mathbf{Q}_{a})^{(t-1-j)})\zeta^{j}\| \nonumber \\
   &\overset{(b)}{\leq} \sum_{j=0}^{t-1}\|\mathbf{J}-(\mathbf{Q}_{r}^C\mathbf{Q}_{a})^{(t-1-j)})\|_{op}\|\zeta^{j}\|\nonumber \\
   &\overset{(c)}{\leq} \sum_{j=0}^{t-1}(\lambda_a\lambda_r^{C})^{(t-1-j)}\|\zeta^j\|,
\end{align}
where $(a)$ uses the (\ref{3}) and (\ref{44}), $(b)$ uses the matrix inequality and $(c)$ uses (\ref{55}).
\begin{align}\label{eq38}
    &\mathbb{E}\|\mathbf{W}^{t}-\mathbf{J}\mathbf{W}^{t}\|^2 \leq \mathbb{E}(\sum_{j=0}^{t-1}(\lambda_a\lambda_r^{C})^\frac{t-1-j}{2}(\lambda_a\lambda_r^{C})^\frac{t-1-j}{2}\|\zeta^{j}\|)^2\nonumber\\
    &\overset{(a)}{\leq} (\sum_{j=0}^{t-1}(\lambda_a\lambda_r^{C})^{t-1-j})(\sum_{j=0}^{t-1}(\lambda_a\lambda_r^{C})^{t-1-j}\mathbb{E}\|\zeta^j\|^2),
\end{align}
where $(a)$ uses the Cauchy inequality.
Obviously, from (\ref{123}) we can get that
\begin{equation}\label{eq39}
    \mathbb{E}\|\zeta^j\|^2 \leq \|\mathbf{Q}_{r}^C\mathbf{Q}_{a}\|^2\cdot\mathbb{E}\|\mathbf{W}^j-\mathbf{Y}^j\|^2 \leq \mathbb{E}\|\mathbf{W}^j-\mathbf{Y}^j\|^2.
\end{equation}

With Lemma 2, we can get
\begin{align}\label{eq40}
        &\mathbb{E}\|\mathbf{W}^j-\mathbf{Y}^j\|^2 \nonumber\\
        &\leq MK(6I\sigma^2+18I^2\zeta^2+18I^2\mathbb{E}\| \nabla f(\mathbf{w}_{mk}^{t}))\|^2)\eta^2.
\end{align}
By combining (\ref{eq38}), (\ref{eq39}), and (\ref{eq40}), we can derive that
\begin{align}
        &\mathbb{E}\|\mathbf{W}^{t}-\mathbf{J}\mathbf{W}^{t}\|^2 \nonumber\\
        &\leq
        \frac{MK(6I\sigma^2+18I^2\zeta^2+18I^2\mathbb{E}\| \nabla f(\mathbf{w}_{mk}^{t})\|^2)\eta^2}{(1-(\lambda_a\lambda_r^{C}))^2}.
\end{align}
Multiplying $\frac{1}{MK}$ on both sides of (41), we can get that
\begin{align}
        &\frac{1}{MK}\mathbb{E}\|\mathbf{W}^{t}-\mathbf{J}\mathbf{W}^{t}\|^2 \nonumber\\
        &\overset{(a)}{=}
        \frac{1}{MK}\sum_{m=1}^{M}\sum_{k=1}^{K}\mathbb{E}\|\mathbf{w}^t_{mk}-\overline{\mathbf{w}^t}\|^2\\
        & \leq
        \frac{(6I\sigma^2+18I^2\zeta^2+18I^2\mathbb{E}\| \nabla f(\mathbf{w}_{mk}^{t})\|^2)\eta^2}{(1-(\lambda_a\lambda_r^{C}))^2},
\end{align}
where $(a)$ uses the fact that $\mathbf{W}^{t}-\mathbf{J}\mathbf{W}^{t} = \left( \begin{array} {c} {{{\mathbf{w}^{t}_{11}-\overline{{{{\mathbf{w}^{t}}}}}}}} \\ {{{\mathbf{w}^{t}_{12}-\overline{{{{\mathbf{w}^{t}}}}}}}} \\ {{\vdots}} \\ {{\mathbf{w}^{t}_{MK}-\overline{{{{\mathbf{w}^{t}}}}}}} \\ \end{array} \right) .
$
This completes the proof.
\section{Proof of Theorem 1}
\textit{Proof:}
Multiplying $\mathbf{Y}^t$ on both sides of (\ref{4}), we can get $\mathbf{JX}^{t+1}=\mathbf{JY}^t$, that is
\begin{equation}
   \overline{{{\mathbf{w}^{t+1}}}}={\overline{{{\mathbf{y}^{t}}}}}. 
\end{equation}

According to the Lipschitz continuity of $\nabla f$ (Assumption 1), we can get that
\begin{align}\label{eq6}
&\mathbb{E}f(\overline{{{\mathbf{w}^{t+1}}}})\leq\mathbb{E}f(\overline{{{\mathbf{w}^{t}}}})\nonumber\\
&\quad+\underbrace{\mathbb{E}\langle\nabla f(\overline{{{\mathbf{w}^{t}}}}),{\overline{{{\mathbf{y}^{t}}}}}-{\overline{{{\mathbf{w}^{t}}}}}\rangle}_{T_1}
+\underbrace{{\frac{L}{2}}\mathbb{E}\|\overline{{{\mathbf{w}^{t+1}}}}-{\overline{{{\mathbf{w}^{t}}}}}\|^{2}}_{T_2}.       
\end{align}

Now we bound the $T_1$ and $T_2$ respectively.
\begin{align}\label{eq44}
    \begin{split}
        T_1&={\mathbb{E}\langle I\nabla f(\overline{{{\mathbf{w}^{t}}}}),({\overline{{{\mathbf{y}^{t}}}}}-{\overline{{{\mathbf{w}^{t}}}}})/I\rangle}\\
        &= \mathbb{E}\langle I\nabla f(\overline{{{\mathbf{w}^{t}}}}), - \eta \nabla f(\overline{{{\mathbf{w}^{t}}}}) + \eta \nabla f(\overline{{{\mathbf{w}^{t}}}}) +({\overline{{{\mathbf{y}^{t}}}}}-{\overline{{{\mathbf{w}^{t}}}}})/I\rangle\\
        &= -\eta I \mathbb{E}\|\nabla f(\overline{{{\mathbf{w}^{t}}}})\|^2 \\
        &\quad+ \mathbb{E}\langle I\nabla f(\overline{{{\mathbf{w}^{t}}}}), \eta \nabla f(\overline{{{\mathbf{w}^{t}}}}) +({\overline{{{\mathbf{y}^{t}}}}}-{\overline{{{\mathbf{w}^{t}}}}})/I\rangle\\
        &= -\eta I\mathbb{E}\|\nabla f(\overline{{{\mathbf{w}^{t}}}})\|^2+ \eta \mathbb{E} \langle \nabla f(\overline{{{\mathbf{w}^{t}}}}),\\
        &\quad\frac{1}{MK}\sum_{m=1}^{M}\sum_{k=1}^{K}\sum_{i=0}^{I-1}(\nabla f_{mk}(\overline{{{\mathbf{w}^{t}}}})-\nabla f_{mk}({{\mathbf{w}_{mk}^{t,i}}},\zeta^i))\rangle\\
        &\leq -\eta I\mathbb{E}\|\nabla f(\overline{{{\mathbf{w}^{t}}}})\|^2 \\
        &\quad+ \eta \mathbb{E} \| \nabla f(\overline{{{\mathbf{w}^{t}}}}) \| \cdot \mathbb{E} \| \sum_{i=0}^{I-1}(\nabla f_{mk}({{\mathbf{w}^{t}_{mk}}})-\nabla f_{mk}({{\mathbf{w}_{mk}^{t,i}}}))\|\\
        &\overset{(a)}{\leq} -\frac{\eta I}{2}\mathbb{E}\|\nabla f(\overline{{{\mathbf{w}^{t}}}})\|^2 \\
        &\quad+ \frac{\eta L^2I}{2}(C_1\eta^2+18I^2\eta^2\frac{\sum_{m=1}^{M}\sum_{k=1}^{K}\mathbb{E}\| \nabla f(\mathbf{w}_{mk}^{t})\|^2}{MK}),\\
    \end{split}
\end{align}
where $(a)$ uses Assumption 1 and Lemma 2.
\begin{align}\label{eq45}
T_2&={\frac{L}{2}}\mathbb{E}\|\overline{{{\mathbf{y}^{t}}}}-{\overline{{{\mathbf{w}^{t}}}}}\|^{2} \nonumber\\
&\leq {\frac{L}{2}\frac{1}{MK}\sum_{m=1}^{M}\sum_{k=1}^{K}\|\mathbf{w}_{mk}^{t,I}-\mathbf{w}_{mk}^{t}\|^2}\nonumber\\
&\overset{(a)}{\leq} {\frac{L}{2}}\eta^2C_1+9LI^2\eta^2\frac{\sum_{m=1}^{M}\sum_{k=1}^{K}\mathbb{E}\| \nabla f(\mathbf{w}_{mk}^{t})\|^2}{MK},
\end{align}
where $(a)$ uses Lemma 2.
\begin{align}\label{eq7}
 &\frac{\sum_{m=1}^{M}\sum_{k=1}^{K}\mathbb{E}\| \nabla f(\mathbf{w}_{mk}^{t})\|^2}{MK}\nonumber\\
 &=\frac{\sum_{m=1}^{M}\sum_{k=1}^{K}\mathbb{E}\| \nabla f(\mathbf{w}_{mk}^{t})-\nabla f(\overline{{{\mathbf{w}^{t}}}})+\nabla f(\overline{{{\mathbf{w}^{t}}}})\|^2}{MK}\nonumber\\
    &\overset{(a)}{\leq} \frac{\sum_{m=1}^{M}\sum_{k=1}^{K}2\mathbb{E}\| \nabla f(\mathbf{w}_{mk}^{t})-\nabla f(\overline{{{\mathbf{w}^{t}}}})\|^2+2\mathbb{E}\|\nabla f(\overline{{{\mathbf{w}^{t}}}})\|^2}{MK}\nonumber \\
    & \leq 2L^2\frac{\sum_{m=1}^{M}\sum_{k=1}^{K}\mathbb{E}\| \mathbf{w}_{mk}^{t}-\overline{{{\mathbf{w}^{t}}}}\|^2}{MK}+2\mathbb{E}\|\nabla f(\overline{{{\mathbf{w}^{t}}}})\|^2\nonumber\\
    &\overset{(b)}{\leq} \frac{2L^2C_2\eta^2}{(1-(\lambda_a\lambda_r^{C}))^2}+2\mathbb{E}\|\nabla f(\overline{{{\mathbf{w}^{t}}}})\|^2,
\end{align}
where $(a)$ uses the inequality $\|\mathbf{a}+\mathbf{b}\|^2\leq 2\|\mathbf{a}\|^2+2\|\mathbf{b}\|^2$ and $(b)$ uses Lemma 3.

Therefore, we can derive from (\ref{eq7}) that
\begin{align}\label{eq47}
        &\frac{\sum_{m=1}^{M}\sum_{k=1}^{K}\mathbb{E}\| \nabla f(\mathbf{w}_{mk}^{t})\|^2}{MK}\nonumber\\
        &\leq \frac{2L^2C_3+2(1-(\lambda_a\lambda_r^{C}))^2\mathbb{E}\|\nabla f(\overline{{{\mathbf{w}^{t}}}})\|^2}{(1-(\lambda_a\lambda_r^{C}))^2-36L^2\eta^2I^2},
\end{align}
where $C_3=6I\eta^2\sigma^2+18I^2\eta^2\zeta^2$.

Thus, by combining (\ref{eq44}), (\ref{eq45}), and (\ref{eq47}), we can represent (\ref{eq6}) as
\begin{equation}\label{eq8}
    \begin{split}
        &\mathbb{E}f(\overline{{{\mathbf{w}^{t+1}}}})\\
&\leq\mathbb{E}f(\overline{{{\mathbf{w}^{t}}}})-\frac{\eta I}{2}\mathbb{E}\|\nabla f(\overline{{{\mathbf{w}^{t}}}})\|^2
+\frac{\eta^3L^2I}{2}C_1+\frac{\eta^2L}{2}C_1\\
&+9\eta^2LI^2(\eta LI+1) \frac{2L^2C_3+2(1-(\lambda_a\lambda_r^{C}))^2\mathbb{E}\|\nabla f(\overline{{{\mathbf{w}^{t}}}})\|^2}{(1-(\lambda_a\lambda_r^{C}))^2-36L^2\eta^2I^2}\\
&\overset{(a)}{\leq} \mathbb{E}f(\overline{{{\mathbf{w}^{t}}}})+(18\eta^3L^2I^3-\frac{\eta I}{2}+18\eta^2LI^2)\mathbb{E}\|\nabla f(\overline{{{\mathbf{w}^{t}}}})\|^2\\
&+\frac{18\eta^3L^4I^3C_3+18\eta^2L^3I^2C_3}{(1-(\lambda_a\lambda_r^{C}))^2}+\frac{\eta^2L}{2}C_1+\frac{\eta^3L^2I}{2}C_1,
    \end{split}
\end{equation}
where $(a)$ needs the stepsize $\eta = \mathcal{O}(\frac{1}{LI\sqrt{T}})$ and $T$ is large.

Reorganize the inequality (\ref{eq8}) and we can have
\begin{align}\label{eq9}
&\mathbb{E}\|\nabla f(\overline{{{\mathbf{w}^{t}}}})\|^2\nonumber\\
        &\overset{(a)}{\leq}  \frac{2f(\overline{\mathbf{w}^t})-2f(\overline{\mathbf{w}^{t+1}})}{(\eta I-36\eta^3L^2I^3-36\eta^2LI^2)}\nonumber\\
&\quad+\frac{\eta^3L^2IC_1+\frac{36\eta^3L^4I^3C_3+36\eta^2L^3I^2C_3}{(1-(\lambda_a\lambda_r^{C}))^2}+\eta^2LC_1}{\eta I-36\eta^3L^2I^3-36\eta^2LI^2},
\end{align}
where $(a)$ needs $\eta I-36\eta^3L^2I^3-36\eta^2LI^2 > 0$. Given the stepsize $\eta = \mathcal{O}(\frac{1}{LI\sqrt{T}})$, we can see that $\eta I-36\eta^3L^2I^3-36\eta^2LI^2 > 0$ as $T$ is large.

Summing the inequality (\ref{eq9}) from $t=0$ to $T-1$, we can obtain the final result as below:
\begin{align}
&\frac{1}{T}\sum_{t=0}^{T - 1}\mathbb{E}\|\nabla f(\overline{{{\mathbf{w}^{t}}}})\|^2 \nonumber\\
&\leq \frac{2f(\overline{\mathbf{w}^0})-2f(\overline{\mathbf{w}^T})}{T(\eta I-36\eta^3L^2I^3-36\eta^2LI^2)}\nonumber\\
&\quad+\frac{\eta^3L^2IC_1+\frac{36\eta^3L^4I^3C_3+36\eta^2L^3I^2C_3}{(1-(\lambda_a\lambda_r^{C}))^2}+\eta^2LC_1}{\eta I-36\eta^3L^2I^3-36\eta^2LI^2}.
\end{align}
This completes the proof. 


\bibliographystyle{IEEEtran}
\bibliography{DFedSat}

\end{document}